\documentclass[aps,pre,subeqn,nofootinbib,notitlepage,amsmath,amsfonts]{revtex4-1}

\usepackage{GlavDesign}

\graphicspath{{Figures/}}

\begin{document}

\title{Transport of heat and mass in a two-phase mixture. From a continuous to a discontinuous description.}

\author{K.~S.~Glavatskiy}
\author{D.~Bedeaux}
\affiliation {
Department of Chemistry, Norwegian University of Science and Technology, NO 7491 Trondheim, Norway.\\
Department of Process and Energy, Technical University of Delft, Leeghwaterstr 44, 2628 CA Delft, The Netherlands. }
\date\today

\begin{abstract}
We present a theory which describes the transport properties of the interfacial region with respect to heat and mass transfer. Postulating the local Gibbs relation
for a continuous description inside the interfacial region, we derive the description of the Gibbs surface in terms of excess densities and fluxes along the
surface. We introduce overall interfacial resistances and conductances as the coefficients in the force-flux relations for the Gibbs surface. We derive relations
between the local resistivities for the continuous description inside the interfacial region and the overall resistances of the surface for transport between the
two phases for a mixture. It is shown that interfacial resistances depend among other things on the enthalpy profile across the interface. Since this variation is
substantial the coupling between heat and mass flow across the surface are also substantial. In particular, the surface puts up much more resistance to the heat and
mass transfer then the homogeneous phases over a distance comparable to the thickness of the surface. This is the case not only for the pure heat conduction and
diffusion but also for the cross effects like thermal diffusion. For the excess fluxes along the surface and the corresponding thermodynamic forces we derive
expressions for excess conductances as integrals over the local conductivities along the surface. We also show that the curvature of the surface affects only the
overall resistances for transport across the surface and not the excess conductivities along the surface.
\end{abstract}

\maketitle


\numberwithin{equation}{section}

\section{Introduction}

\label{sec/Introduction}

During evaporation and condensation heat and mass are transported through the interface. The common description of these phenomena uses assumptions which are
questionable. For instance, one usually assumes equilibrium conditions at the interface \cite{Krishna/MassTransfer}. Furthermore one neglects coupling effects
between heat and mass transfer \cite{Krishna/MassTransfer}. The first one is a zeroth order approximation in deviations from equilibrium and it is desirable to
extend it to a more accurate theory. This is, for instance, done in the monograph by Kjelstrup and Bedeaux \cite{kjelstrupbedeaux/heterogeneous}. Neglecting
coupling coefficients was shown to violate the second law of thermodynamics for a one-component system \cite{Bedeaux2005}. Coupling is important since the
corresponding resistances for transport across the surface depend on the enthalpy of vaporization. The significance of this quantity in this context is due to the
large difference between the liquid and the vapor values of the enthalpy. Neglecting the coupling coefficients has been shown to imply that the enthalpy of
vaporization is zero \cite{kjelstrupbedeaux/heterogeneous}, which is clearly incorrect.

The situation becomes even more complicated when one considers evaporation in mixtures. These processes happen in such industrial applications as distillation and
therefore the precise description is important. Depending on conditions one can get mass fluxes of components in the same direction or in the direction opposite to
the corresponding thermodynamic forces.

There has been done a number of studies of the interfacial transport for one- and two-component systems: experiments \cite{Fang1999a, Fang1999c, Ward2001,
Badam2007, Mills2002, James2006}, molecular dynamic simulations \cite{Rosjorde2000, Rosjorde2001, Kjelstrup2002, Simon2004, jialin/longrange} , kinetic theory
\cite{Pao1971a, Sone1973, Cipolla1974, Bedeaux1990, Bedeaux1992} and square gradient continuous description \cite{bedeaux/vdW/I, bedeaux/vdW/II, bedeaux/vdW/III,
bedeaux/vdW/IV, glav/grad1, glav/grad2}. All these works use different approaches, which allows one to investigate different aspects. Most studies were for
one-component systems. It was verified that non-equilibrium thermodynamics could be used to describe the data by Fang and Ward \cite{Fang1999a, Fang1999c,
Bedeaux1999b}, see also \cite{Badam2007, kjelstrupbedeaux/heterogeneous}. The description of transport through surfaces, using non-equilibrium thermodynamics
\cite{kjelstrupbedeaux/heterogeneous}, gives a unifying treatment also for the results of the molecular dynamics simulations, kinetic theory and the square gradient
theory, both for single component systems and mixtures. An alternative approach to interpret the data is the statistical rate theory \cite{Fang1999b, Ward1999}. See
for a critical discussion of the description of evaporation and condensation the paper by Bond and Struchtrup \cite{Struchtrup2004}.

The results of the molecular dynamics analysis as well as from the square gradient theory give continuous stationary state profiles of, for instance, the
temperature and the densities through the surface. Both descriptions \cite{Rosjorde2000, Simon2004, jialin/longrange, bedeaux/vdW/II, glav/grad2} conclude that the
surface is in local equilibrium. This shows that the surface as described by the excess densities introduced by Gibbs \cite{Gibbs/ScientificPapers} is a separate
thermodynamic system. This makes the systematic description of the surface using non-equilibrium thermodynamics given in \cite{kjelstrupbedeaux/heterogeneous}
possible. One of the points of interest is the dependence of the resulting overall interfacial resistances on the continuous profiles. Once we have a description
which relates the total resistances of the Gibbs surface to the continuous profiles of, in particular, the local resistivities and the enthalpy profiles, we can
study which aspects of the continuous description are most important..

It is the aim of this paper to obtain the general discrete approach to describe transport properties through and along the surface on the basis of the continuous
description. As the interfacial region is thin, it is natural to consider the whole region as a single entity. In equilibrium, the common method to do this is to
introduce the properties of the Gibbs surface as excess densities. Away from equililibrium the surface should be described in terms of excess densities and fluxes.
To obtain these one needs the actual profiles of the thermodynamic densities and fluxes as well as the profiles extrapolated from the homogeneous regions. Within
such a description, an equilibrium surface may be considered as an autonomous phase \cite{Rowlinson1982}. Given the validity of local equilibrium it is possible to
speak about the Gibbs surface also in non-equilibrium states, as was mentioned above. In this paper we will establish the link between the non-equilibrium
continuous description and the transport properties of the whole surface both through and along the surface directly. Crucial in the analysis is that the surface
thickness is very small compared to length scales over which variables change in the homogeneous phases and along the surface. For fluids this leads to
2-dimensional isotropy of the surface discussed in \cite{glav/grad1}.

We study the transport through the surface in curvilinear coordinates. This allows us to consider both transport through planar interfaces and a phenomenon like
nucleation. The curvature of the surface plays an important role in nucleation. The agreement of the classical theory \cite{Becker1935, Abraham1974, Feder1966} with
experiments is in some cases rather poor \cite{Fladerer2006, Iland2007, Wedekind2007}. The work in this paper will make it possible to analyze effects due to
curvature on the overall resistances for transport of heat and mass through the surface of the nucleus in more detail.

One of the issues this paper focuses on is the overall resistances of the surface to heat and mass transfer through the surface. The thermodynamic forces, which
lead to heat and mass transfer through the surface, are given by the differences between the values of the inverse temperature and chemical potentials divided by
the temperature in gas and liquid phase. It is known that there are jumps of these quantities across the surface \cite{Fang1999a, Fang1999c, Ward2001, Badam2007,
Mills2002, James2006}. These jumps are given in terms of the total heat and mass fluxes using these overall resistances. We refer to a monograph by Kjelstrup et al.
\cite{kjelstrupbedeaux/heterogeneous} for a systematic derivation of these linear relations in the context of non-equilibrium thermodynamics of surfaces. The other
issue studied in this paper is the excess conductances along the surface. Also for these conductances integral relations are derived, expressing them in the
behavior of the continuous conductivities in the interfacial region.

The analysis in this paper will only consider stationary states. The resulting overall resistances and conductances can also be used to analyze non-stationary
behavior. We will neglect viscous contributions to the pressure tensor.

From the calculation of the overall surface resistances we are able to understand the aspects in the continuous description which are most important for the size of
the overall resistances. These are found to be the nature of the local resistivity profiles and the enthalpy profiles across the surface. Both can be obtained from
the particular continuous description used. From molecular dynamics simulations there is evidence that local resistivities have a peak \cite{Simon2004, surfres} in
the interfacial region. Within the context of the square gradient model studied in \cite{bedeaux/vdW/I, bedeaux/vdW/II, bedeaux/vdW/III, bedeaux/vdW/IV, glav/grad1,
glav/grad2} this peak was modeled by a square gradient contribution. Such a peak clearly contributes significantly to excess resistances. The enthalpy is important
because the value of the enthalpy is significantly different in gas and liquid phase. The overall resistances are found to contain integrals over products of the
local resistivity profiles and the enthalpy profiles. The location and the nature of the changeover from a large positive enthalpy in the vapor to a large negative
enthalpy in the liquid relative to the change over and the peak in the local resistivities is important for the size and the sign of the overall resistances. The
overall conductances are found to contain the excesses of the conductivities for the continuous description.

In \secr{sec/Thermodynamics} we give a short discussion of the continuous description of the interface using non-equilibrium thermodynamics. We use the Gibbs
relation valid for mixtures in the interfacial region found in the context of the square gradient theory, discussed in \cite{glav/grad1, glav/grad2}, which we
believe to be general. This enables us to obtain the rate of entropy production everywhere in the interfacial region. In \secr{sec/Excess} we define the excess
densities needed to describe the non-equilibrium Gibbs surface. Furthermore, we show how the small thickness of the interfacial region allows us to split all
phenomena in separate contributions parallel and perpendicular to the surface in \secr{sec/Excess/Surface}. In \secr{sec/Integral/Accross} we derive integral
relations for the overall interfacial resistances in terms of the local resistivities and the equilibrium enthalpy profiles. In \secr{sec/Integral/Along} we derive
the integral relations for the overall interfacial conductances. We further derive the corresponding relations for the coefficients for the measurable heat fluxes
in \secr{sec/Measurable}. This is done in two steps, for normal resistances in \ssecr{sec/Measurable/Perpendicular} and for parallel conductances in
\ssecr{sec/Measurable/Parallel}. In \secr{sec/Discussion} we discuss the consequences of the derived relations. In particular we discuss the influence of the
surface geometry and the enthalpy profile across the interfacial region. Finally, in \secr{sec/Conclusions} we give concluding remarks.

\section{Thermodynamics of the interfacial region}\label{sec/Thermodynamics}

\subsection{Gibbs relation}

Consider a two phase $n$-component mixture. Let $T$ be the temperature and $ p $ be a scalar pressure. Let $h_{i}$ be the partial enthalpy of the $i$-th component,
$\mu_{i}$ be the chemical potential and $\mu_{in}\equiv \mu_{i}-\mu_{n}$ be the chemical potential difference of the $i$-th component with the $n$-th component.
Furthermore, let $h$, $u$, $s$, $v$ be the mass specific enthalpy, internal energy, entropy and volume respectively, $\rho \equiv 1/v$ be the overall mass density
and $\xi_{i}\equiv \rho_{i}/\rho $ be the mass fraction of the $i$-th component. All these fields depend on the position $\vR$\ and the time $t$. For a planar
interface $p(\vRt)$ is equal to the parallel pressure $p_{\parallel }({\mathbf{ r}},\,t)$.

The interfacial region is known to have a large density variation over a distance of the order of a nanometer. This was the reason for van der Waals
\cite{Waals1893, Waals1979} to assume, that, for instance, the Helmholtz energy density of the surface depends not only on the local properties, but also on their
gradients. Due to this dependence the surface tension is unequal to zero. For a two-phase system it becomes an additional thermodynamic characteristics and has to
be accounted in the description of the surface. In the continuous description the role of the surface tension is played by the tension tensor $\gamma_{\alpha \beta
}(\vRt)$. For a planar surface in equilibrium the only nonzero element of this tension tensor is the $xx$ element along the diagonal which equals
$p_{\perp}(\vR)-p_{\parallel }(\vR)$. The integral of this difference over the $x$-coordinate normal to the surface gives the surface tension of the planar surface.
For a curved surface the tension tensor is no longer diagonal. For a fluid-fluid interface it remains symmetric, however. For the relation with the surface tension
in that case we refer to \cite{glav/grad1, glav/grad2}.

A systematic non-equilibrium thermodynamic description requires the Gibbs relation expressing the rate of change of the entropy density in those of the other
thermodynamic quantities. In \cite{glav/grad1, glav/grad2}\ we used the following form of the Gibbs relation
\begin{equation}
T(\vRt)\,\frac{ds(\vRt)}{dt}=\frac{du(\vRt)}{dt}-\sum_{i=1}^{n-1}{\mu_{in}(\vRt)\frac{d\xi_{i}(\vR,\,t)}{dt}}+p(\vRt)\frac{dv(\vRt)}{dt}-v(\vRt)\,{\mathrm{v}}_{\beta
}(\vRt)\frac{\partial \gamma _{\alpha \beta }(\vRt)}{\partial x_{\alpha }} \label{eq/Entropy/Gibbs/01}
\end{equation}
where $\vvelocity$ is the barycentric velocity, and $d/dt=\partial /\partial t+\mathrm{v}_{\alpha }\partial /\partial x_{\alpha }=\partial /\partial
t+\vvelocity\spd\nabla$ is the substantial time derivative. Furthermore we use the summation convention over double Greek indices. Compared to the form of the Gibbs
relation used in the homogeneous regions \cite{deGrootMazur}, it has an additional contribution proportional to the divergence of the tension tensor $\gamma
_{\alpha \beta }$. This contribution is only nonzero in the interfacial region. The use of it emphasizes the difference between the surface and the homogeneous
regions. The above expression is equivalent to the one given in \cite{bedeaux/vdW/I}\ for the one-component two phase fluid. In \cite{glav/grad1, glav/grad2}\ we
motivate the choice of the new term in the interfacial region. This motivation is not a derivation, however, and the validity of the above Gibbs relation is
therefore postulated. The justification of the specific choice of the Gibbs relation is on the one hand a comparison with experiments and on the other hand the
thermodynamic consistency of the results. Such a consistancy was found to be lacking for other choices.

The next step in the thermodynamic description is to give the relation between the thermodynamic potentials. In a bulk region they are known to be homogeneous
functions of the first order. It is not obvious that this property remains valid inside the interfacial region. It follows however from \eqr{eq/Entropy/Gibbs/01}
that the internal energy density has to be a homogeneous function of the first order of the entropy density, volume and mass densities. Indeed, all the terms in
\eqr{eq/Entropy/Gibbs/01} but the last one have the common form. The last term is proportional to the gradient of the tension tensor, which is not an extensive
quantity, however, so it does not contribute to the internal energy. We therefore may write the following relation both for the homogeneous as well for the
interfacial regions
\begin{equation}
u(\vRt)=\mu_{n}(\vRt)+\sum_{i=1}^{n-1}{\mu_{in}(\vRt)\,\xi_{i}(\vRt)}-p(\vRt)\,v(\vR,\,t)+T(\vRt)\,s(\vRt) \label{eq/Entropy/Gibbs/03}
\end{equation}
We emphasize the role of the interfacial region here. While the thermodynamic potentials are related via the standard thermodynamic relations, the Gibbs relation in
the interfacial region is different from the bulk Gibbs relation. Important is that the standard thermodynamic relations contain $p(\vRt)$, which is equal to the
parallel pressure for a planar interface and which is very different in the interfacial region from the asymptotic values away from the surface.

Substituting \eqr{eq/Entropy/Gibbs/03} into \eqr{eq/Entropy/Gibbs/01} we obtain
\begin{equation}
s\,\frac{dT}{dt}+\frac{d\mu_{n}}{dt}+\sum_{i=1}^{n-1}{\xi_{i}\frac{d\mu _{in}}{dt}}-v\,\frac{dp}{dt}-v\,{\mathrm{v}}_{\beta }\frac{\partial \gamma _{\alpha \beta
}}{\partial x_{\alpha }}=0  \label{eq/Entropy/Gibbs/04}
\end{equation}
This is the Gibbs-Duhem equation for a two-phase multi-component mixture. Note that since $\mu_{in}\equiv \mu_{i}-\mu_{n}$ and $\sum_{i=1}^{n}{\xi _{i}=1}$ we have
\begin{equation}
\frac{d\mu_{n}}{dt}+\sum_{i=1}^{n-1}{\xi_{i}\frac{d\mu_{in}}{dt}} =\sum_{i=1}^{n}{\xi_{i}\frac{d\mu_{i}}{dt}}  \label{eq/Entropy/Gibbs/05}
\end{equation}
which is the usual contribution to the Gibbs-Duhem equation associated with the chemical potentials. To simplify the notation we will not usually indicate the
$\vRt$ dependence.

For a stationary state the derivative $\partial /\partial t$ gives zero and \eqr{eq/Entropy/Gibbs/04} takes the following form
\begin{equation}
{\mathrm{v}}_{\beta }\,\left( s\,\frac{\partial T}{\partial x_{\beta }}+ \frac{\partial \mu_{n}}{\partial x_{\beta }}+\sum_{i=1}^{n-1}{\xi_{i}\frac{
\partial \mu_{in}}{\partial x_{\beta }}}-v\,\frac{\partial \sigma_{\alpha
\beta }}{\partial x_{\alpha }}\right) =0  \label{eq/Entropy/Gibbs/04a}
\end{equation}
where $\sigma_{\alpha \beta }=p\,\delta_{\alpha \beta }+\gamma_{\alpha \beta }$ is the thermodynamic pressure tensor.

The equation of motion for a stationary state takes the following form:
\begin{equation}
\rho {\mathrm{v}}_{\alpha }\frac{\partial \mathrm{v}_{\beta }}{\partial x_{\alpha }}=-\frac{\partial \sigma_{\alpha \beta }}{\partial x_{\alpha }} -\rho
{\frac{\partial (\vg\spd\vR)}{\partial x_{\beta } }}
\end{equation}
where $\vg\spd\vR$\ is the gravitational potential. Contracting this equation with $\mathrm{v}_{\beta }$\ and dividing by $\rho $ \ the result may be written as:
\begin{equation}
{\mathrm{v}}_{\beta }\,\left( v\,\frac{\partial \sigma_{\alpha \beta }}{
\partial x_{\alpha }}+\frac{\partial ({\mathrm{v}}^{2}/2-\vg\!\cdot
\!\vR)}{\partial x_{\beta }}\right) =0  \label{eq/Entropy/Gibbs/04b}
\end{equation}
Using it together with \eqr{eq/Entropy/Gibbs/04a} we get
\begin{equation}
\vvelocity\spd\sum_{i=1}^{n}{\xi_{i}\left(\nabla\frac{\wmu_{i}}{T} - \wh_{i}\nabla\frac{1}{T}\right)}=0 \label{eq/ExcessEntropy/04a}
\end{equation}
which is the Gibbs-Duhem equation valid both in the homogeneous regions as well as in the interfacial region in a stationary state. Here $\widetilde{ \mu
}_{i}\equiv \mu_{i}+{\mathrm{v}}^{2}/2-\!\vg\!\cdot {\mathbf{r} }$ and $\wh_{i}\equiv h_{i}+{\mathrm{v}}^{2}/2-\vg\!\cdot \!\vR$. Note, that $\wh_{i}\equiv
\widetilde{\mu }_{i}+Ts_{i}$, where $s_{i}$ is the partial entropy.

\subsection{Entropy production}\label{sec/Entropy/Balance}

The entropy balance equation is
\begin{equation}
\rho \,\frac{ds}{dt}=-\nabla \spd\vJ_{s}+\sigma_{s} \label{eq/Entropy/Balance/01}
\end{equation}
with the entropy flux $\vJ_{s}\equiv \vJ_{s,tot}-\rho s{ \mathbf{v}}$ and the entropy production $\sigma_{s}$. Given the Gibbs relation \eqr{eq/Entropy/Gibbs/01} it
is shown in \cite{glav/grad1, glav/grad2} that the entropy flux and the entropy production are given by the following expressions
\begin{subequations}\label{eq/Entropy/Balance/02}
\begin{equation}
\vJ_{s}=\displaystyle\frac{1}{T}\left( \vJ_{q}-\sum_{i=1}^{n-1}{\mu_{in}\,\vJ_{i}}\right) \label{eq/Entropy/Balance/02a}
\end{equation}
\begin{equation}
\sigma_{s}=\displaystyle\vJ_{q}\spd\nabla \frac{1}{T} -\sum_{i=1}^{n-1}{\vJ_{i}\spd\nabla \frac{\mu_{in}}{T}} \label{eq/Entropy/Balance/02b}
\end{equation}
\end{subequations}
where $\vJ_{q}$ and $\vJ_{i}$ are the heat and diffusion fluxes
\begin{equation}
\begin{array}{rl}
\vJ_{q} & \equiv \displaystyle\vJ_{e}-\rho \,\vvelocity \,e-p\,\vvelocity=\vJ_{e}-\vJ_{m}\,(h+{\mathrm{v}}^{2}/2-
\vg\spd\vR) \\
&  \\
\vJ_{i} & \equiv \rho_{i}\,(\vvelocity_{i}-\vvelocity)=\vJ_{\xi_{i}}-\xi_{i}\,\vJ_{m}
\end{array}
\label{eq/Entropy/Balance/03}
\end{equation}
The diffusive mass fluxes satisfy the relation $\sum_{i=1}^{n}{\vJ_{i}}=0$. In turn, the energy flux $\vJ_{e}$ and the mass fluxes $\vJ_{\xi_{i}}\equiv
\rho_{i}\,\vvelocity_{i}$ and $\vJ_{m}\equiv \rho \,\vvelocity$ are convenient quantities since in stationary states
\begin{equation}
\nabla \spd\vJ_{e}=0,\quad \nabla \spd\vJ_{\xi _{i}}=0,\quad \nabla \spd\vJ_{m}=0 \label{eq/Entropy/Balance/03a}
\end{equation}
Furthermore, it follows from \eqr{eq/Entropy/Balance/01} that in a stationary state
\begin{equation}
\sigma_{s}=\nabla \spd\vJ_{s,tot}=\nabla \spd{\mathbf{J }}_{s}+\rho \,\vvelocity\spd\nabla s=\nabla \spd(\vJ_{s}+ \rho\,s\vvelocity) \label{eq/Entropy/Balance/04}
\end{equation}
Using the Gibbs-Duhem equation \eqref{eq/Entropy/Gibbs/04a} and the conservation laws, under stationary state conditions, it is possible to show that
\begin{equation}
\sigma_{s}=\vJ_{e}\spd\nabla \frac{1}{T}-\sum_{i=1}^{n}{\vJ_{\xi_{i}}\spd\nabla \frac{\widetilde{\mu_{i}}}{T}} \label{eq/Entropy/Balance/05}
\end{equation}

The fluid-fluid interface has a two-dimensional isotropy. A detailed discussion of what this means in a continuous 3-dimensional description is given in
\cite{glav/grad1}.\ The essential element in this discussion is the fact that the interfacial region is thin compared to the radius of curvature, $\delta \ll \ell
$. This implies that the 3-dimensional vectorial fluxes and forces in the entropy production \eqref{eq/Entropy/Balance/05} split up into 2-dimensional vectorial
components parallel to the surface and scalar components normal to the surface. We may therefore write
\begin{equation}
\sigma_{s}=\sigma_{s,\parallel }+\sigma_{s,\perp }  \label{eq/Entropy/Balance/05a}
\end{equation}
where the parallel and normal contributions are
\begin{eqnarray}
\sigma_{s,\parallel } &=& \vJ_{e,\parallel}\spd\nabla_{\parallel}\frac{1}{T}-\sum_{i=1}^{n}{\vJ_{\xi_{i},\parallel}\spd\nabla_{\parallel}\frac{\widetilde{\mu_{i}}}{T}}
\label{eq/Entropy/Balance/05b} \\
\sigma_{s,\perp } &=& J_{e,\perp}\nabla_{\perp}\frac{1}{T} - \sum_{i=1}^{n}{J_{\xi_{i},\perp}\nabla_{\perp}{\frac{\wmu_{i}}{T}}} \label{eq/Entropy/Balance/05c}
\end{eqnarray}
Two-dimensional isotropy now implies that the parallel and the normal forces and fluxes do not couple and the force-flux relations for parallel and perpendicular
contribution are independent. As we will see below, it is convenient to write these relations in terms of conductivities for parallel contributions and in terms of
resistivities for perpendicular contributions. We therefore have the following force-flux relations in the parallel direction
\begin{equation}\label{eq/Resistivities/01a}
\begin{array}{rl}
\vJ_{e,\parallel} =&\displaystyle  \ell_{qq,\parallel }^{e}\,\nabla_{\parallel}\frac{1}{T} + \sum_{i=1}^{n}{\ell_{qi,\parallel }^{e}\,\left(-\nabla_{\parallel
}\frac{\wmu_{i}}{T}\right)}
\\
\vJ_{\xi _{j},\parallel } =&\displaystyle  \ell_{iq,\parallel }^{e}\,\nabla_{\parallel }\frac{1}{T} +
\sum_{j=1}^{n}{\ell_{ij,\parallel}^{e}\,\left(-\nabla_{\parallel}\frac{\wmu_{i}}{T}\right)}
\end{array}
\end{equation}
and in the normal direction
\begin{equation}\label{eq/Resistivities/01}
\begin{array}{rl}
\displaystyle\nabla_{{\perp }}\frac{1}{T} & =\displaystyle r_{qq,\perp }^{e}\,J_{e,{\perp }}+\sum_{i=1}^{n}{r_{qi,{\perp }}^{e}\,J_{\xi_{i},{\perp }}}
\\
-\displaystyle\nabla_{{\perp }}\frac{\widetilde{\mu }_{j}}{T} & = \displaystyle r_{jq,{\perp }}^{e}\,J_{e,{\perp }}+\sum_{i=1}^{n}{r_{ji,{\perp }
}^{e}\,J_{\xi_{i},{\perp }}}
\end{array}
\end{equation}
The 2-dimensional isotropy implies that transport coefficients for both contributions (conductivities for parallel contribution as well as resistivities for normal
contribution) are scalar. The off-diagonal coefficients of both sets of force-flux equations satisfy the Onsager reciprocal relations. Inside the interfacial region
the normal resistivities are different from the corresponding parallel resistivities. Away from the interfacial region the fluid has a 3-dimensional isotropy and
the corresponding resistivities (or conductivities) for the parallel and the normal direction become equal. In order to be consistent with the sign conventions used
in the monograph by Kjelstrup et al. \cite{kjelstrupbedeaux/heterogeneous} we changed the sign convention we used in \cite{glav/grad1, glav/grad2}. For this purpose
we changed the sign of the chemical forces which results in a change of the sign of the cross coefficients between the energy flux and the mass fluxes.

We emphasize the reader's attention on the following difference in the convention. It is common to give the constitutive force-flux relations for $n$-component
fluid in terms of $n$ independent fluxes, e.g. the heat flux $\vJ_{q}$ and $n-1$ diffusive fluxes $\vJ_{i}$. Those fluxes are linearly independent and the
corresponding matrix of transport coefficients (resistivities or conductivities) is non-singular. If one uses $n+1$ fluxes or forces as variables, the corresponding
matrix of the transport coefficients has $n+1$ rows and columns and is therefore singular. This makes it impossible to convert resistivities into conductivities
that easily, so the corresponding representation should be fixed. The use of particular representation (resistivities or conductivities) is dictated by the surface
symmetry and will be apparent later. The singularity of the matrix of transport coefficients does not however restrict us much. For instance, there is one-to-one
correspondence between the coefficients of $n$-squared matrix and $n+1$-squared matrix (see e.g. \appr{sec/Appendix/Resistivities}). Furthermore, the singularity of
the transport coefficients matrix does not affect the Onsager reciprocal relations, cf. \cite[ch.VI, \S 3]{deGrootMazur}. We can therefore operate these
coefficients normally, unless we want to invert the matrix.

\section{The Gibbs surface}\label{sec/Excess}

The Gibbs surface is described with the help of excesses of different quantities in the interfacial region. In equilibrium only excesses of the densities are used
in the description. Non-equilibrium involves many other quantities which vary through the interfacial region and which therefore require a proper definition of the
corresponding excess quantity. In this section we will define the \textit{excess densities} 
in non-equilibrium. In the following sections the definition of the \textit{excess resistance}, the \textit{excess conductance}, and the \textit{excess fluxes} will
be given.

Because of the variety in nature of the quantities we want take the excess of, we shall use the unified notation for the excess. Namely, we will use
$\Oexcess_{\tau}[q]$ to denote the excess of a profile $q(\vR)$, where the subscript $\tau$ indicates the nature of the the quantity $q$. For example, $\OexcessD$
will indicate the excess of a density and $\OexcessJ$ will indicate the excess of a flux. Even though both $\OexcessD$ and $\OexcessJ$ represent the property of the
Gibbs surface, they are defined differently.

The definition of an excess requires the normal direction ${\vn}$ to be defined in the interfacial region. The surface may be curved and we may introduce
curvilinear orthogonal coordinates $\vR\equiv (x_{1},x_{2},x_{3})$ with $x_{1}$ being the normal coordinate and ${\vR}_{\parallel}\equiv (x_{2},x_{3})$ being the
tangential coordinates. For the stationary states these coordinates are independent of the time.

Let $\phi $ be a variable defined in the interfacial and the homogeneous regions. Furthermore, for any variable $\phi $, let $\phi ^{b}$, where superscript $b$
stands either for $\ell $ or for $g$, be the variable $\phi $ extrapolated from the bulk to the surface region. The extrapolation is done using the description in
homogeneous phases. Outside of the interfacial region $\phi ^{b}$ and $\phi $ are almost identical but inside the surface $ \phi ^{b}$ in general differs from $\phi
$. In order to make this comparison more precise, let $x^{g,s}(\vR_{\parallel })$ and $x^{\ell ,s}(\vR_{\parallel })$ be the boundaries of the interfacial region at
the gas and liquid side respectively. These boundaries are chosen such that the extrapolated value $\phi ^{b}$\ differs some small fraction (like for instance 0.1
promille or 0.01 promille) from the actual value $\phi $ along these boundaries. Given the small size of this difference we can use the following identity for the
extrapolated variables
\begin{equation}
\phi^{b}(\xsb(\vR_{\parallel}),\vR_{\parallel}) = \phi(\xsb(\vR_{\parallel}),\vR_{\parallel }) \label{eq/Excess/Definition/01}
\end{equation}

The location of the boundaries of the interfacial region depend on the variable $\phi $\ and the accuracy used. For a sensible choice of the variable and the
accuracy, all choices are equivalent. In contrast to equilibrium, in non-equilibrium description all the bulk variables vary in space and so do the extrapolated
quantities. This is true both for densities, like the density $\rho $, which vary only slightly in the bulk phases, and for intensive quantities, like the
temperature, which may vary significantly even in the bulk phases. If one extrapolates any function $F$ of a number of variables the result equals the original
function of the extrapolated variables:
\begin{equation}
F^{b}(\ldots ,\phi ,\ldots) = F(\ldots ,\phi^{b},\ldots ) \label{eq/Excess/Definition/02}
\end{equation}
This is a consequence of the fact that the equality is obvious away from the surface. For instance, $(\mu_{i}/T)^{g} = \mu_{i}^{g}/T^{g}$.

We note however, that even though we introduce the extrapolated variables in such a way that \eqr{eq/Excess/Definition/01} and \eqr{eq/Excess/Definition/02} are
true, the use of any numerical procedure to obtain the extrapolated variables will introduce small errors. This is true in particular for non-equilibrium when the
extrapolated variables are not constant. This is however a technical question, not a fundamental one. Usually the extrapolation involves polynomials in order to fit
an actual curve, which introduces a non-zero error in the extrapolated curve.

On the basis of some criterium one chooses a dividing surface in the interfacial region. The location along the ${x_{1}}$\ coordinate is denoted by
${\xs(\vR_{\parallel })}$. This location is usually, though not necessarily, between $x^{g,s}(\vR_{\parallel })$ and $x^{\ell ,s}(\vR_{\parallel })$. We will
discuss below which criteria are used for this choice.

\subsection{Excess density}\label{sec/Excess/Definition}

If $\phi (\vR)$ is a density per unit of volume in the 3-dimensional space, we can define its excess $\excessD{\phi}$ as \cite{albano/sing}
\begin{equation}
\excessD{\phi}(\xsRll,\vRll) \equiv \frac{1}{\lame_{2}^{s}\,\lame_{3}^{s}}\,\int_{\xgsb}^{\xlsb}{dx_{1}\,\lame_{1}\,\lame_{2}\,\lame_{3}\,\phi^{ex}(\vR; \xsRll))}
\label{eq/Excess/Definition/03}
\end{equation}
where
\begin{equation}
\phi^{ex}(\vR;\xsRll) \equiv \phi(\vR)-\phi^{g}(\vR)\,\Theta(\xsRll-x_{1}) - \phi^{\ell}(\vR)\,\Theta(x_{1}-\xsRll) \label{eq/Excess/Definition/04}
\end{equation}
Furthermore, $\lame_{\alpha}\equiv \lame_{\alpha}(x_{1},\vRll)$ are Lame coefficients for the curvilinear coordinates and $ \lame_{\alpha}^{s}{(\vRll)} \equiv
\lame_{\alpha}(\xsRll, \vRll)$ for $i=2,3$ are Lame coefficients for the curvilinear coordinates along the dividing surface. Given that $\phi(\vR)$ is a density per
unit of volume, the excess $\excessD{\phi}(\xsRll, \vRll)$ is a density per unit of surface. The excess depends on the position of the dividing surface $\xsRll$.
For ease of notation we will not further indicate the dependence of $\xs$, $\xgsb$ and $\xlsb$ on $\vR_{\parallel}$ explicitly.


The key quantity in the non-equilibrium description is the entropy production. As for any density, one can introduce the excess entropy production for the Gibbs
surface. It will determine the dissipation rate of the whole surface. Consider the entropy production given in \eqr{eq/Entropy/Balance/05}. All the terms have the
form $\vJ\spd\nabla \varpi $, where $\vJ$ is a flux and $\varpi $ is some scalar function. According to \eqr{eq/Entropy/Balance/03a} $\nabla \spd{\vJ }=0$ in the
stationary states, so $\vJ\spd\nabla\varpi = \nabla\spd(\vJ\varpi)$. We show in \appr{sec/Appendix/Excess/Gradient/3d} that this leads to
\begin{equation}
\excessD{\vJ\spd\nabla \varpi} = (J_{\perp}\varpi)^{\ell} - (J_{\perp}\varpi)^{g} + \excessD{\nabla_{\parallel}\spd({\vJ}_{\parallel}\varpi)}
\label{eq/Excess/Surface/01}
\end{equation}
where all the functions on the right hand side are evaluated at $\vR ^{\scriptstyle s}$. We see, that the excess entropy production of the Gibbs surface in the
stationary states can be split into two contributions: the one takes into account only perturbations perpendicular to the surface, the other depends only on the
perturbations parallel to the surface:
\begin{equation}\label{eq/Excess/Surface/01a}
\excessD{\sigma_{s}} = \excessD{\sigma_{s, \parallel}} + \excessD{\sigma_{s, \perp}}
\end{equation}
We note, that such a separation does not require the surface to be two-dimensionally isotropic.

\section{A thin interface}\label{sec/Excess/Surface}

Substantial simplification in the expression for the excess entropy production can be made if one consider typical conditions, when the interfacial region is thin.
As was discussed in \cite{glav/grad1}, the interfacial region breaks the 3-dimensional isotropy of the system. In addition to a typical macroscopic size of the
problem $\dalong$, there exists the microscopic size $\dthrough \ll \dalong$, the surface width, which is of the order of few nanometers.

First, we shall make a note on the definition of the excess density. For a thin surface the Lame coefficients $\lame_{\alpha} = \lame_{\alpha}(x_{\perp},
\vR_{\parallel})$ do not vary much along $x_{\perp}$ coordinate. To a relative order $\left(\dthrough / \dalong \right)$ they are equal to $\lame^{s}_{\alpha}$:
\begin{equation}\label{eq/Excess/Surface/12}
\lame_{\alpha}(x_{\perp}, \vR_{\parallel}) = \lame_{\alpha}(\xs, \vR_{\parallel}) + \dthrough\,\frac{d\lame_{\alpha}}{dx_{\perp}} + \ldots
\end{equation}
Substituting this expansion into \eqr{eq/Excess/Definition/03}, we get
\begin{equation}\label{eq/Excess/Definition/03a}
\excessD{\phi} = \excess{\phi} + O\left(\dthrough / \dalong\right)
\end{equation}
where
\begin{equation}\label{eq/Resistivities/03}
\excess{\phi} \equiv \displaystyle \int_{\xgsb}^{\xlsb}{dx_{1}\lame_{1}\,\phi^{ex}(\vR; \xs)}
\end{equation}

For each flux in \eqr{eq/Entropy/Balance/03a} we can write $\nabla _{\perp }J_{\perp }+\nabla_{\parallel }\spd\vJ_{\parallel }=0 $. This gives an approximate
relation for the order of magnitude
\begin{equation}
\frac{|\Delta_{\perp }J_{\perp }|}{\Delta_{\perp} x_{\perp }} \simeq \frac{|\Delta _{\parallel }J_{\parallel }|}{\Delta_{\parallel} x_{\parallel }}
\label{eq/Excess/Surface/04}
\end{equation}
where $\Delta_{\perp }$ and $\Delta_{\parallel }$ denote a typical change in the corresponding direction. There are quantities which change drastically over the
distances of the order $\dthrough$ in the direction perpendicular to the surface, so $\Delta_{\perp} x_{\perp } \simeq \dthrough$. However, a significant change of
any quantity along the surface may happen only on a length scale $\dalong$, which is of the order of either the radii of curvature or the system size, so
$\Delta_{\parallel} x_{\parallel } \simeq \dalong$. Because of this property of a surface, we may not expect that $|\Delta_{\perp }J_{\perp }|\ll |\Delta
_{\parallel }J_{\parallel }|$, i.e. that the change of the parallel component of a flux on a macroscopic scale along the surface is much larger then the change of
the perpendicular component of that flux on a microscopic scale through the surface. For the fluxes for which changes $|\Delta_{\perp }J_{\perp }|$ and
$|\Delta_{\parallel }J_{\parallel }|$ are of the same order of magnitude $\Delta J$, \eqr{eq/Excess/Surface/04} takes the form $\Delta J/\dthrough \approx \Delta
J/\dalong $, which can hold only if $ \Delta J=0$, since $\dthrough \ll \dalong$. This means that both $\Delta_{\perp }J_{\perp }=0$ and $\Delta_{\parallel
}J_{\parallel }=0$. If $|\Delta _{\perp }J_{\perp }|\gg |\Delta_{\parallel }J_{\parallel }|$ this statement becomes even stronger.
We then may require that for a thin surface\footnote{ For the special case of a system with planar surface in cartesian coordinated with all the fluxes directed
perpendicular to the surface, these equations follow straightforwardly.}
\begin{subequations}\label{eq/Excess/Surface/05}
\begin{equation}
\nabla_{\perp }J_{\perp }(\vR)=0  \label{eq/Excess/Surface/05a}
\end{equation}
\vspace{-2.015pc}
\begin{equation}
\nabla_{\parallel }\spd\vJ_{\parallel }(\vR)=0 \label{eq/Excess/Surface/05b}
\end{equation}
\end{subequations}
Thus, for a surface with a thickness much smaller than the radii of curvature, the stationary state condition $\nabla \spd\vJ=0$ has the form of
\eqr{eq/Excess/Surface/05} in an interfacial region.

The extrapolated fluxes $\vJ^{b}$ satisfy the same equations
\begin{subequations}\label{eq/Excess/Surface/07}
\begin{equation}
\nabla_{\perp }J_{\perp }^{b}(\vR)=0  \label{eq/Excess/Surface/07a}
\end{equation}
\vspace{-2.0075pc}
\begin{equation}
\nabla_{\parallel }\spd\vJ_{\parallel }^{b}(\vR)=0 \label{eq/Excess/Surface/07b}
\end{equation}
\end{subequations}
since the extrapolated flux fields also satisfy $\nabla \spd{\vJ }^{b}=0$.

Consider now \eqr{eq/Excess/Surface/05a} and \eqr{eq/Excess/Surface/07a}. Both of them are first order ordinary differential equations, the solutions of which
contains additive constants. These constants must be the same, since according to \eqr{eq/Excess/Definition/01} $J_{\perp }^{b}(\vR^{b,s})=J_{\perp }(\vR^{b,s})$ at
the boundary points. It follows that $J_{\perp }^{b}$ and $J_{\perp }$ are the same functions:
\begin{equation}\label{eq/Excess/Surface/02a}
J_{\perp }^{g}(\vR)=J_{\perp }^{\ell }(\vR)=J_{\perp }(\vR)
\end{equation}
Note that \eqr{eq/Excess/Surface/02a} does not lead to the relation $ J_{\perp }(x^{g,s},\vR_{\parallel }) = J_{\perp }(x^{\ell ,s},\vR_{\parallel })$.
\eqr{eq/Excess/Surface/02a} is the relation between values of different functions at the same point but not the relation between values of the same function at
different points. However, it follows from \eqr{eq/Excess/Surface/05a} that in curvilinear coordinates $\partial (\lame_{2}\,\lame_{3}\,J_{\perp })/(\partial
x_{\perp })=0$ and therefore
\begin{equation}\label{eq/Excess/Surface/02b}
\lame_{2}(x_{\perp})\,\lame_{3}(x_{\perp})\,J_{\perp }(x_{\perp}) = const = \lame_{2}^{s}\,\lame_{3}^{s}\,J(\xs)
\end{equation}
but not $J_{\perp }=const$.

Consider now \eqr{eq/Excess/Surface/05b} and \eqr{eq/Excess/Surface/07b}. It follows that
\begin{equation}\label{eq/Excess/Surface/08}
\begin{array}{lr}
\nabla_{\parallel }\spd(\vJ_{\parallel}\varpi) &= \vJ_{\parallel}\spd\nabla_{\parallel}\varpi  \\\\
\nabla_{\parallel }\spd(\vJ_{\parallel}^{b}\varpi) &= \vJ_{\parallel}^{b}\spd\nabla_{\parallel}\varpi
\end{array}
\end{equation}
Applying both \eqr{eq/Excess/Surface/02a} and \eqr{eq/Excess/Surface/08} to \eqr{eq/Excess/Surface/01} we get
\begin{equation}\label{eq/Excess/Surface/09}
\excess{\vJ\spd\nabla \varpi }=J_{\perp}(\varpi^{\ell }-\varpi^{g}) + \excess{\vJ_{\parallel}\spd\nabla_{\parallel}\varpi}
\end{equation}
Applying \eqr{eq/Excess/Surface/09} to each term in \eqr{eq/Entropy/Balance/05} for the entropy production we obtain the general form of the excess entropy
production for a surface in stationary state
\begin{equation}
\begin{array}{rl}
\excess{\sigma_{s}} = & \displaystyle J_{e,\,\perp }\left( \frac{1}{T^{\ell }}-\frac{1}{T^{g}} \right) -\sum_{i=1}^{n}{J_{\xi_{i},\,\perp }\left(
\frac{\widetilde{\mu } _{i}^{\ell }}{T^{\ell }}-\frac{\widetilde{\mu }_{i}^{g}}{T^{g}}\right) }\\
&  \\
& + \displaystyle\excess{\vJ_{e,\,\parallel}\spd\nabla_{\parallel }\frac{1}{T}} -
\sum_{i=1}^{n}{\excess{\vJ_{\xi_{i},\,\parallel}\spd\nabla_{\parallel}\frac{\wmu_{i}}{T}}} \\
\end{array}
\label{eq/ExcessEntropy/01}
\end{equation}

We see, that for a thin surface, the excess entropy production splits into independent contributions which are caused by forces perpendicular and parallel to the
surface. This property has the same origin as the 2-dimensional isotropy of the surface \cite{glav/grad1}.

It is desired to have the excess entropy production in the form $\excess{\sigma_{s}} = \sum{J_{j}\,X_{j}}$, i.e. as a sum of independent contributions, each of
which can is a product of a flux and a force. However, as one can see from \eqr{eq/ExcessEntropy/01} the parallel contributions do not have such a form. Further
information about the symmetry of a system is required. We will therefore consider the cases when the forces are applied perpendicular to the surface and parallel
to the surface separately.

\section{Integral relations for transport through the surface}\label{sec/Integral/Accross}

Consider forces applied only in the direction perpendicular to the surface. Then $\sigma_{s,\parallel} = 0$ and the local entropy production $\sigma_{s} =
\sigma_{s,\perp}$ which is given by \eqr{eq/Entropy/Balance/05c}. The corresponding force-flux relation are given by \eqr{eq/Resistivities/01}.

The excess entropy production given by \eqr{eq/ExcessEntropy/01} simplifies to the following
\begin{equation}
\excess{\sigma_{s}} = \displaystyle J_{e,\,\perp }\left( \frac{1}{T^{\ell }}-\frac{1}{T^{g}} \right) -\sum_{i=1}^{n}{J_{\xi_{i},\,\perp }\left(
\frac{\widetilde{\mu } _{i}^{\ell }}{T^{\ell }}-\frac{\widetilde{\mu }_{i}^{g}}{T^{g}}\right) }\\
\label{eq/ExcessEntropy/01c}
\end{equation}
The fluxes in \eqr{eq/ExcessEntropy/01c} are evaluated at the dividing surface $\xs$. Furthermore, $T^{\ell }\equiv T^{\ell }(\xs) $ and $T^{g}\equiv T^{g}(\xs)$
are the temperatures extrapolated from the liquid and gas to the dividing surface. The analogous meaning have $\wmu_{i}^{\ell}$ and $\wmu_{i}^{g}$.

Following the traditional procedure in non-equilibrium thermodynamics, we write the force-flux relations based on the form of the excess entropy production
\eqref{eq/ExcessEntropy/01c} as:
\begin{equation}\label{eq/Resistivities/02}
\begin{array}{rl}
\displaystyle\frac{1}{T^{\ell }}-\frac{1}{T^{g}} & =\displaystyle R_{qq,\perp}^{e}\,J_{e,\perp }+\sum_{i=1}^{n}{R_{qi,\perp }^{e}\,J_{\xi_{i},\perp }} \\
-\displaystyle\left( \frac{\widetilde{\mu }_{j}^{\ell }}{T^{\ell }}-\frac{ \widetilde{\mu }_{j}^{g}}{T^{g}}\right) & =\displaystyle R_{jq,\perp }^{e}\,J_{e,\perp
}+\sum_{i=1}^{n}{R_{ji,\perp }^{e}\,J_{\xi_{i},\perp }}
\end{array}
\end{equation}

We now compare \eqr{eq/Resistivities/02} with \eqr{eq/Resistivities/01}. \eqr{eq/Resistivities/02} for the force-flux relations has the following form
\begin{equation}\label{eq/Resistivities/12}
\varpi^{\ell}(\xs)-\varpi^{g}(\xs)= R_{\perp}^{e}(\xs)\,J_{\perp}(\xs)
\end{equation}
The resistance $R_{\perp}^{e}$ in general depends on the choice of the dividing surface. \eqr{eq/Resistivities/01} has the following form
\begin{equation}\label{eq/Resistivities/11}
\nabla_{\perp}\varpi(x) = r_{\perp}^{e}(x)\,J_{\perp}(x)
\end{equation}
Let us apply $\Oexcess$ to both sides of \eqr{eq/Resistivities/11}. We note, that we use the excess operator $\Oexcess$, given by \eqr{eq/Resistivities/03} exactly,
not any other operator which is equal to $\Oexcess$ only to the order $\dthrough / \dalong$. Using \eqr{eq/Excess/Surface/02b} and the fact that in curvilinear
coordinates $\nabla_{\perp} = \lame_{1}^{-1}\,\partial/\partial{x}_{1}$, one obtains
\begin{equation}\label{eq/Resistivities/13}
\varpi^{\ell}(\xs) - \varpi^{g}(\xs) = \excessR{r_{\perp}}(\xs)\,J_{\perp}(\xs)
\end{equation}
where we introduced the \textit{excess resistance} operator $\OexcessR$. For a resistivity profile $r(\vR)$ it is defined as
\begin{equation}\label{eq/Resistivities/14}
\excessR{r} \equiv \lame_{2}^{s}\,\lame_{3}^{s}\int_{\xgsb}^{\xlsb}{dx_{1}\frac{\lame_{1}}{\lame_{2}\lame_{3}}\,r^{ex}(\vR;\xs)}
\end{equation}
Comparing \eqr{eq/Resistivities/13} with \eqr{eq/Resistivities/12} we therefore conclude that
\begin{equation}\label{eq/Resistivities/16}
R_{\perp}^{e} = \excessR{r_{\perp}^{e}}
\end{equation}
This is the general form of the resistivity integral relation for the transport normal to the surface.

Given that the forces are applied only in the direction perpendicular to the surface, \eqr{eq/Resistivities/16} is valid for the interfacial region of any
thickness. We now make use of the fact that the surface is thin. For such a case we expand the Lame coefficients $\lame_{\alpha}(x)$ around $\xs$ and obtain
\begin{equation}\label{eq/Resistivities/15}
\excessR{r} = \excess{r} + O\left(\dthrough / \dalong\right)
\end{equation}
so that to a relative order $\left(\dthrough / \dalong\right)$ the excess resistance $\excessR{r}$ is given by $\excess{r}$.

Applying the above procedure to each term in \eqr{eq/Resistivities/01} and \eqr{eq/Resistivities/02} we may conclude that
\begin{equation}\label{eq/Resistivities/17}
\begin{array}{rl}
R_{qq,\perp }^{e} & = \excess{r_{qq,\perp }^{e}} \\
&  \\
R_{qi,\perp }^{e} & = \excess{r_{qi,\perp }^{e}} \\
&  \\
R_{ji,\perp }^{e} & = \excess{r_{ji,\perp }^{e}} \\
&
\end{array}
\end{equation}
to the order $\left(\dthrough / \dalong\right)$. \eqr{eq/Resistivities/15} gives integral relations for the resistivity coefficients for the case that one uses the
total energy flux to describe the transport through the surface. They show that resistivities normal to the surface are additive, as one would expect.

\section{Integral relations for transport along the surface}\label{sec/Integral/Along}

Consider now the perturbations applied only in the direction parallel to the surface. Then $\sigma_{s,\perp} = 0$ and the local entropy production $\sigma_{s} =
\sigma_{s,\parallel}$ which is given by \eqr{eq/Entropy/Balance/05b}. The corresponding force-flux relation are given by \eqr{eq/Resistivities/01a}.

\subsection{Excess entropy production}

The excess entropy production given by \eqr{eq/ExcessEntropy/01} simplifies to the following
\begin{equation}
\excess{\sigma_{s,\parallel}} = \displaystyle\excess{\vJ_{e,\,\parallel}\spd\nabla_{\parallel }\frac{1}{T}} -
\sum_{i=1}^{n}{\excess{\vJ_{\xi_{i},\,\parallel}\spd\nabla_{\parallel}\frac{\wmu_{i}}{T}}} \\
\label{eq/ExcessEntropy/01b}
\end{equation}
This form of the excess entropy production does not allow us to write the constitutive force flux relations yet. However, the fact that there are no perturbations
in the direction perpendicular to the surface, allows us to simplify this expression further.

Each term in \eqr{eq/ExcessEntropy/01b} has a form $\excess{\vJ_{\parallel}\spd\nabla_{\parallel }\varpi}$, where the possible perturbation is controlled by a
scalar function $\varpi(x_{\perp}, \vR_{\parallel})$. For perturbations applied only in the direction parallel to the surface, $\varpi$ is independent of
$x_{\perp}$. The same argument is applicable to the extrapolated property $\varpi^{b}$. We show in \appr{sec/Appendix/Excess/Gradient/2d} that this leads to the
following relation
\begin{equation}\label{eq/Excess/Surface/06}
\excessD{\vJ_{\parallel}\spd\nabla_{\parallel}\varpi} = \excessJ{\vJ_{\parallel}}\spd\nabla_{\parallel}\varpi
\end{equation}
where we have introduced the \textit{excess flux} operator $\OexcessJ$. For a parallel component $\vJ_{\parallel} \equiv (J_{2}, J_{3})$ of a flux $\vJ \equiv
(J_{\perp },\vJ_{\parallel})$, the excess parallel flux \cite{albano/sing} is defined as $\excessJ{\vJ_{\parallel}}\equiv (\excessJy{J_{2}},\excessJz{J_{3}})$,
where
\begin{equation}
\begin{array}{rl}
\excessJy{J_{2}}(\xs,\vR_{\parallel }) & \equiv \displaystyle\frac{1}{\lame_{3}^{s}}\,\int_{\xgsb}^{ \xlsb}{dx_{1}\,\lame_{1}\,\lame
_{3}\,J_{2}^{ex}(\vR;\xs)} \\
\excessJz{J_{3}}(\xs,\vR_{\parallel }) & \equiv \displaystyle\frac{1}{\lame_{2}^{s}}\,\int_{\xgsb}^{ \xlsb}{dx_{1}\,\lame_{1}\,\lame _{2}\,J_{3}^{ex}(\vR;\xs)}
\end{array}
\label{eq/Excess/Definition/05}
\end{equation}
where
\begin{equation}
\vJ^{ex}(\vR;\xs)\equiv \vJ(\vR)-\vJ^{g}(\vR)\,\Theta (\xs-x_{1})-\vJ^{\ell }(\vR)\,\Theta (x_{1}-\xs) \label{eq/Excess/Definition/06}
\end{equation}
As discussed in more detail in \cite{albano/sing} possible excess fluxes normal to the surface, calculated in a reference frame that moves along with the surface in
the normal direction, play no role in the discreet description of transport through the surface. Here we note that for the special case of stationary states the
surface is not moving in the normal direction.

For a thin surface one can expand the Lame coefficients around $\xs$ and obtain to a relative order $\left(\dthrough / \dalong \right)$ that
\begin{equation}\label{eq/Excess/Definition/05a}
\excessJ{\vJ_{\parallel}} = \excess{\vJ_{\parallel}} + O\left(\dthrough / \dalong\right)
\end{equation}
where operator $\Oexcess$ for a flux $\vJ_{\parallel}$ is still given by \eqr{eq/Resistivities/03}.

Following \cite{albano/sing} we denote the excess parallel flux $\excessJ{\vJ_{\parallel}}$ as $\vJ_{\parallel}^{s}$.

Consider now $\nabla_{\parallel}\varpi$ which appears in \eqr{eq/Excess/Surface/06}. In \cite{glav/grad2} we discussed how to define the temperature $T^{s}$ and the
chemical potentials $\mu_{i}^{s}$ of the surface in terms of the surface tension and the relative adsorptions. These quantities are independent of the choice of the
dividing surface when the surface is planar. For a curved surface differences of the relative order of the distance between the dividing surfaces divided by the
curvature $\left(\dthrough / \dalong \right) $ may appear. The quantity $\varpi $ considered above can be written as a function of the temperature and the chemical
potentials alone. We can therefore evaluate this quantity at the temperature $T^{s}$ and the chemical potentials $\mu_{i}^{s}$ of the surface. This then gives $
\varpi^{s}$, which therefore will be independent of the choice of the dividing surface to a relative order $\dthrough / \dalong$. As we have discussed, the
variation of $\varpi$ along the surface is the same as the variation of $\varpi^{b}$ along the surface. It therefore follows that the variation of $\varpi^{s}$
along the surface should also be the same to this order. \eqr{eq/Excess/Surface/06} takes the following form
\begin{equation}\label{eq/Excess/Surface/11}
\excess{\vJ_{\parallel}\spd\nabla_{\parallel}\varpi} = \vJ_{\parallel}^{s}\spd\nabla _{\parallel }\varpi^{s}
\end{equation}
to this order.

Furthermore, we identify the excess entropy production in \eqr{eq/ExcessEntropy/01b} as the surface entropy production: $\sigma_{s,\parallel}^{s}$. With this
notation, substituting \eqr{eq/Excess/Surface/11} into \eqr{eq/ExcessEntropy/01b} and taking into account the terms to the order $\dthrough / \dalong$ we get for
the excess entropy production
\begin{equation}
\sigma_{s,\parallel}^{s} = \vJ_{e,\parallel}^{s}\spd\nabla_{\parallel}\frac{1}{T^{s}} -
\sum_{i=1}^{n}{\vJ_{\xi_{i},\parallel}^{s}\spd\nabla_{\parallel}\frac{\widetilde{\mu_{i}}^{s}}{T^{s}}} \label{eq/ExcessEntropy/02a}
\end{equation}
which has the form used in \cite{bedeaux/advchemphys}.

\subsection{Integral relations}

According to the traditional procedure in non-equilibrium thermodynamics, the force-flux relations which follow from the excess entropy production
\eqref{eq/ExcessEntropy/02a} are
\begin{equation}\label{eq/Resistivities/02a}
\begin{array}{rl}
\vJ_{e,\parallel}^{s} =& \displaystyle L_{qq,\parallel}^{e}\,\nabla_{\parallel}\frac{1}{T^s} + \sum_{i=1}^{n}{L_{qi,\parallel }^{e}\,\left(-\nabla_{\parallel
}\frac{\wmu_{i}^s}{T^s}\right)}
\\\\
\vJ_{\xi_{j},\parallel}^{s} =& \displaystyle L_{iq,\parallel }^{e}\,\nabla_{\parallel}\frac{1}{T^s} +
\sum_{j=1}^{n}{L_{ij,\parallel}^{e}\,\left(-\nabla_{\parallel}\frac{\wmu_{i}^s}{T^s}\right)}
\end{array}
\end{equation}
Due to the isotropy of the surface the conductance coefficients $L_{\parallel}^{e}$ are scalars.

We now compare \eqr{eq/Resistivities/02a} with \eqr{eq/Resistivities/01a}. \eqr{eq/Resistivities/02a} for the force-flux relations have the following form
\begin{equation}\label{eq/Resistivities/22}
\vJ_{\parallel}^{s}(\xs) = L_{\parallel}^{e}(\xs)\,\nabla_{\parallel}\varpi^{s}
\end{equation}
Both $L_{\parallel}^{e}$ and $\excess{\vJ_{\parallel}}$ in general depend on the choice of the dividing surface, while $\nabla_{\parallel}\varpi^{s}$ does not.
\eqr{eq/Resistivities/01a} has the following form
\begin{equation}\label{eq/Resistivities/21}
\vJ_{\parallel}(x) = \ell_{\parallel}^{e}(x)\,\nabla_{\parallel}\varpi(x)
\end{equation}

Let us apply $\OexcessJ$ to the both sides of \eqr{eq/Resistivities/21}. We note, that even though the excess operator $\OexcessJ$ given by
\eqr{eq/Excess/Definition/05} is equal to the operator $\Oexcess$ to the order $\dthrough / \dalong$, we will first keep the higher order terms. Direct substitution
yields
\begin{equation}\label{eq/Resistivities/23}
\vJ_{\parallel}^{s}(\xs) = \excessL{\ell_{\parallel}^{e}}(\xs)\spd\nabla_{\parallel}\varpi^{s}
\end{equation}
where we introduced the \textit{excess conductance} operator $\OexcessL$. For a conductivity profile $\ell(\vR)$ it is defined as a matrix
\begin{equation}\label{eq/Resistivities/24}
\excessL{\ell} \equiv
\begin{pmatrix}
\displaystyle \frac{\lame_{2}^{s}}{\lame_{3}^{s}}\,\int{dx_{1}\lame_{1}\frac{\lame_{3}}{\lame_{2}}\,\,\ell^{ex}} & 0 \\
0 & \displaystyle \frac{\lame_{3}^{s}}{\lame_{2}^{s}}\,\int{dx_{1}\lame_{1}\frac{\lame_{2}}{\lame_{3}}\,\,\ell^{ex}}
\end{pmatrix}
\end{equation}

Next we would like to compare \eqr{eq/Resistivities/22} with \eqr{eq/Resistivities/23} in order to relate the interfacial conductance $ L_{\parallel}^{e}$ to the
excess conductance $\excessL{\ell_{\parallel}^{e}}$. This however is not possible directly, since in general the former one is a scalar while the latter one is a
matrix. However, if we take into account that the surface is thin, we expand the Lame coefficients $\lame_{\alpha}(x)$ around $\xs$ and obtain
\begin{equation}\label{eq/Resistivities/25}
\excessL{\ell} \equiv
\begin{pmatrix}
\excess{\ell} & 0 \\
0 & \excess{\ell}
\end{pmatrix} + O\left(\dthrough / \dalong\right)
\end{equation}
so that $\excessL{\ell} = \excess{\ell}$ to a relative order $\left(\dthrough / \dalong\right)$. We note, that to this order the excess $\excessL{\ell}$ becomes
scalar. Comparing then \eqr{eq/Resistivities/23} with \eqr{eq/Resistivities/22} we can conclude that
\begin{equation}\label{eq/Resistivities/26}
L_{\parallel}^{e} = \excess{\ell_{\parallel}^{e}}
\end{equation}
to the order $\left(\dthrough / \dalong\right)$. Neglecting curvature effects, this is the general form of the conductivity integral relation for the transport
parallel to the surface.

Comparing \eqr{eq/Resistivities/01a} with \eqr{eq/Resistivities/02a} and using \eqr{eq/Resistivities/26} we may conclude that
\begin{equation}\label{eq/Resistivities/27}
\begin{array}{rl}
L_{qq,\parallel }^{e} & = \excess{\ell_{qq,\parallel }^{e}} \\
&  \\
L_{qi,\parallel }^{e} & = \excess{\ell_{qi,\parallel }^{e}} \\
&  \\
L_{ji,\parallel }^{e} & = \excess{\ell_{ji,\parallel }^{e}} \\
\end{array}
\end{equation}
\eqr{eq/Resistivities/15} represents integral relations for the conductivity coefficients for the case that one uses the total energy flux to describe the transport
along the surface. They show that conductivities parallel to the surface are additive, as one would expect.

\section{Measurable heat fluxes}\label{sec/Measurable}
\subsection{Local transport coefficients}

It is convenient to write the local entropy production also in terms of the measurable heat flux $\vJ_{q}^{\prime }$, which is defined as
\begin{equation}
\vJ_{q}^{\prime }\equiv \vJ_{q}-\sum_{i=1}^{n}{h_{i}\vJ_{i}}=\vJ_{e}-\sum_{i=1}^{n}{\wh_{i}\,{\vJ }_{\xi_{i}}}  \label{eq/ExcessEntropy/03}
\end{equation}
In terms of measurable heat fluxes the expression for the entropy production becomes
\begin{equation}
\sigma_{s}=\vJ_{q}^{\,\prime}\spd\,\nabla \frac{1}{T}-\sum_{i=1}^{n-1} {\vJ_{i}\spd\,\left( \nabla \frac{\mu_{in}}{T}-h_{in}\,\nabla \frac{1}{ T}\right) }
\label{eq/ExcessEntropy/06}
\end{equation}
where $h_{in} \equiv h_{i}-h_{n}$. \eqr{eq/Entropy/Balance/04}, \eqr{eq/Entropy/Balance/05} and \eqr{eq/ExcessEntropy/06}, which give the entropy production in
stationary states, are equivalent expressions. To second order in the deviation from equilibrium one may, and we will, use the equilibrium values of $h_{in}$ in
\eqr{eq/ExcessEntropy/06}. The resulting expression for the entropy production is then only equivalent to the other two to this order.

As in \secr{sec/Entropy/Balance} we can split the local entropy production into parallel and perpendicular contributions based on the two-dimensional isotropy of
the interfacial region, see \eqr{eq/Entropy/Balance/05a}, so that
\begin{eqnarray}
\sigma_{s,\parallel } &=& \vJ_{q,\parallel}^{\,\prime}\spd\nabla_{\parallel}\frac{1}{T} - \sum_{i=1}^{n-1}{\vJ_{i,\parallel}\spd
\left(\nabla_{\parallel}\frac{{\mu}_{in}}{T} - h_{in,eq}\nabla_{\parallel}\frac{1}{T} \right)} \label{eq/ExcessEntropy/05b}
\\
\sigma_{s,\perp } &=& J_{q,\,\perp}^{\,\prime}\nabla_{\perp}\frac{1}{T} - \sum_{i=1}^{n-1}{J_{i,\perp}\left(\nabla_{\perp}\frac{\mu_{in}}{T} -
h_{in,\,eq}\nabla_{\perp}\frac{1}{T}\right)} \label{eq/ExcessEntropy/05c}
\end{eqnarray}
This leads to the following force-flux relations in the parallel direction
\begin{equation}\label{eq/Measurable/04a}
\begin{array}{rl}
\vJ_{q,\parallel}^{\,\prime} = &\displaystyle  \ell_{qq}^{\,\prime}\,\nabla_{\parallel}\frac{1}{T} +
\sum_{i=1}^{n-1}{\ell_{qi}^{\,\prime}\,\left(-\nabla_{\parallel}\frac{{\mu}_{in}}{T}+h_{in,eq}\nabla_{\parallel}\frac{1}{T} \right)}
\\\\
\vJ_{j,\parallel} = &\displaystyle \ell_{iq}^{\,\prime}\,\nabla_{\parallel }\frac{1}{T} +
\sum_{i=1}^{n-1}{\ell_{ij}^{\,\prime}\,\left(-\nabla_{\parallel}\frac{{\mu}_{in}}{T}+h_{in,eq}\nabla_{\parallel}\frac{1}{T} \right)}
\end{array}
\end{equation}
and in the normal direction
\begin{equation}
\begin{array}{rl}
\displaystyle\nabla_{\perp} \frac{1}{T} & = \displaystyle r_{qq}^{\,\prime}\,J_{q,\,\perp}^{\,\prime} + \sum_{i=1}^{n-1}{r_{qi}^{\,\prime}\,J_{i,\,\perp}}
\\\\
-\displaystyle\nabla_{\perp} \frac{\mu_{in}}{T} + h_{in,\,eq}\nabla_{\perp}\frac{1}{T} & = \displaystyle r_{jq}^{\,\prime}\,J_{q,\,\perp}^{\,\prime } +
\sum_{i=1}^{n-1}{r_{ji}^{\,\prime}\,J_{i,\,\perp}}
\end{array}
\label{eq/Measurable/04}
\end{equation}
Further on, for the transport through the surface we will use only resistivities and for the transport along the surface we will use only conductivities. We have
therefore omitted subscripts $\perp$ and $\parallel$ for corresponding transport coefficients to simplify the notation.

We now relate the local resistivities associated with the measurable heat flux to the local resistivities associated with the total energy flux. Comparing\footnote{
The details of this procedure are given in \appr{sec/Appendix/Resistivities}.} \eqr{eq/Measurable/04} with \eqr{eq/Resistivities/01} we obtain
\begin{equation}  \label{eq/Measurable/06}
\begin{array}{rll}
r^{e}_{qq} = & \displaystyle r_{qq}
\\\\
r^{e}_{qi} = & \displaystyle r_{qi} - r_{qq}\,\wh_{i,eq}, & i=\nrange
\\\\
r^{e}_{ji} = & \displaystyle r_{ji} - r_{qi}\,\wh_{j,eq} - r_{jq}\,\wh_{i,eq} + r_{qq}\,\wh_{j,eq}\,\wh_{i,eq}, & i,j=\nrange\\
\end{array}
\end{equation}
where $r_{qq}$, $r_{qi}$, and $r_{ji}$ are defined as
\begin{equation}  \label{eq/Measurable/06a}
\begin{array}{rl}
r_{qq} \equiv & r^{\,\prime}_{qq}
\\\\
r_{qi} \equiv &
\begin{cases}
\displaystyle - \sum_{k=1}^{n-1}{r^{\,\prime}_{qk}\,\xi_{k}} + r^{\,\prime}_{qi}, & i=\nmorange \\
\displaystyle - \sum_{k=1}^{n-1}{r^{\,\prime}_{qk}\,\xi_{k}}, & i=n
\end{cases}
\\\\
r_{ji} \equiv &
\begin{cases}
\displaystyle \sum_{k=1}^{n-1}\sum_{l=1}^{n-1}{r^{\,\prime}_{kl}\,\xi_{k}\, \xi_{l}} - \sum_{k=1}^{n-1}{\xi_{k}\,(r^{\,\prime}_{ki}+\xi_{k}\,r^{\,\prime}_{jk})} +
r^{\,\prime}_{ji}, & i,j=\overline{1,n\!-\!1}
\\
\displaystyle \sum_{k=1}^{n-1}\sum_{l=1}^{n-1}{r^{\,\prime}_{kl}\,\xi_{k}\, \xi_{l}} - \sum_{k=1}^{n-1}{\xi_{k}\,r^{\,\prime}_{jk}}, & i=n,\, j=
\overline{1,n\!-\!1} \\
\displaystyle \sum_{k=1}^{n-1}\sum_{l=1}^{n-1}{r^{\,\prime}_{kl}\,\xi_{k}\, \xi_{l}} - \sum_{k=1}^{n-1}{\xi_{k}\,r^{\,\prime}_{ki}}, & i=\overline{
1,n\!-\!1},\, j=n \\
\displaystyle \sum_{k=1}^{n-1}\sum_{l=1}^{n-1}{r^{\,\prime}_{kl}\,\xi_{k}\, \xi_{l}}, & i,j=n
\end{cases}
\end{array}
\end{equation}

We now relate the local conductivities. Comparing\footnote{The details of this procedure are given in \appr{sec/Appendix/Conductivities}} \eqr{eq/Resistivities/01a}
and \eqr{eq/Measurable/04a} we get
\begin{equation}  \label{eq/ExcessEntropy/13}
\begin{array}{rll}
\ell^{e}_{qq} = & \displaystyle \ell_{qq} + \sum_{j=1}^{n-1}{h_{jn,eq}\,(\ell_{qj}+\ell_{jq})} + \sum_{j=1}^{n-1}\sum_{i=1}^{n-1}{h_{jn,eq}\,h_{in,eq}\,\ell_{ji}}
\\\\
\ell^{e}_{qi} = & \displaystyle \ell_{qi} + \sum_{j=1}^{n-1}{h_{jn,eq}\,\ell_{ij}}, & i=\nrange
\\\\
\ell^{e}_{ji} = & \displaystyle \ell_{ji},  & i,j=\nrange\
\end{array}
\end{equation}
where $\ell_{qq}$, $\ell_{qi}$, and $\ell_{ji}$ are defined as
\begin{equation}  \label{eq/ExcessEntropy/14}
\begin{array}{rl}
\ell_{qq} \equiv  &\ell^{\,\prime}_{qq}
\\\\
\ell_{qi} \equiv &%
\left\{\begin{array}{rll}
\displaystyle &\ell^{\,\prime}_{qi}, & i=\nmorange \\
\displaystyle - \sum_{k=1}^{n-1} &\ell^{\,\prime}_{qk}, & i=n %
\end {array}\right.
\\\\
\ell_{ji} \equiv & %
\left\{\begin{array}{rll}
\displaystyle &\ell^{\,\prime}_{ji}, & i,j=\nmorange\\
\displaystyle - \sum_{k=1}^{n-1} &\ell^{\,\prime}_{jk}, & i=n,\, j=\nmorange \\
\displaystyle - \sum_{k=1}^{n-1} &\ell^{\,\prime}_{ki}, & i=\nmorange,\, j=n \\
\displaystyle \sum_{k=1}^{n-1}\sum_{l=1}^{n-1}  &\ell^{\,\prime}_{kl}, & i,j=n%
\end{array}\right.
\end{array}
\end{equation}

In the rest of the section we will derive integral relations for the measurable resistivities and conductivities. As such, we will relate the measurable interfacial
resistances to the measurable resistivity profiles and the measurable interfacial conductance to the measurable conductivity profiles. The reason for this is that
the coefficients which correspond to the measurable heat flux can be obtained experimentally.

\subsection{Transport through the surface}\label{sec/Measurable/Perpendicular}

The excess entropy production due to transport through the surface can be written in terms of measurable heat fluxes extrapolated from either gas of the liquid
phase tho the surface. The extrapolated measurable heat flux is defined similarly to \eqr{eq/ExcessEntropy/03}:
\begin{equation}
J_{q,\perp}^{\,\prime ,\,b} = J_{e,\perp} - \sum_{i=1}^{n}{\wh_{i}^{b}\,J_{\xi_{i},\perp}}  \label{eq/ExcessEntropy/03a}
\end{equation}
In equilibrium the fluxes are equal to zero. To linear order in these fluxes we may replace the enthalpies by their equilibrium values $\wh_{i,\,eq} \equiv h_{i,eq}
- \vg\spd\rs$. In the following equations one should therefore use $\wh_{i,\,eq}$ in stead of $\wh_{i}$. This is similar to the use of the equilibrium enthalpy
difference profiles in \eqr{eq/ExcessEntropy/06}. Using \eqr{eq/ExcessEntropy/03a}, the excess entropy production \eqref{eq/ExcessEntropy/01c} becomes to linear
order
\begin{equation}\label{eq/ExcessEntropy/10a}
\excess{\sigma_{s,\perp}} = J_{q,\perp}^{\,\prime,\,g} \left(\frac{1}{T^{\ell}}-\frac{1}{T^{g}}\right) - \sum_{i=1}^{n}{J_{\xi_{i},\perp}\left[\left(
\frac{\wmu_{i}^{\ell}}{T^{\ell}}-\frac{\wmu_{i}^{g}}{T^{g}}\right) - \wh_{i,eq}^{g}\left(\frac{1}{T^{\ell}}-\frac{1}{T^{g}}\right)\right]}
\end{equation}
The measurable heat flux $J_{q,\perp }^{\,\prime ,\,g}$ is the one calculated for the gas side of the surface. The similar expression can be written for the
measurable flux $J_{q,\perp }^{\,\prime ,\,\ell}$ calculated for the liquid side of the surface.

The corresponding force-flux equations for the gas-side are
\begin{equation}
\begin{array}{rl}
\displaystyle\frac{1}{T^{\ell }}-\frac{1}{T^{g}} & =\displaystyle
R_{qq}^{\,\prime\,g}\,J_{q,\,\perp}^{\,\prime,\,g}+\sum_{i=1}^{n}{R_{qi}^{\,\prime\,g}\,J_{\xi_{i},\,\perp}}
\\\\
-\displaystyle\left(\frac{\wmu_{i}^{\ell}}{T^{\ell}} - \frac{\wmu_{i}^{g}}{T^{g}}\right) + \wh_{i,eq}^{g}\left(\frac{1}{T^{\ell}}-\frac{1}{T^{g}}\right) &
=\displaystyle R_{jq}^{\,\prime\,g}\,J_{q,\,\perp}^{\,\prime,\,g} + \sum_{i=1}^{n}{R_{ji}^{\,\prime\,g}\,J_{\xi_{i},\,\perp}}
\end{array}
\label{eq/Measurable/01}
\end{equation}

Comparing \eqr{eq/Resistivities/02} with \eqr{eq/Measurable/01} we find
\begin{equation}
\begin{array}{rl}
R_{qq}^{\,\prime\,g} & =R_{qq}^{e} \\
&  \\
R_{qi}^{\,\prime\,g} & =R_{qi}^{e} + \wh_{i,\,eq}^{g}\,R_{qq}^{e} \\
&  \\
R_{ji}^{\,\prime\,g} & =R_{ji}^{e} + \wh_{i,\,eq}^{g}\,R_{jq}^{e}+\wh_{j,\,eq}^{g}\,R_{qi}^{e} + \wh_{i,\,eq}^{g}\,\wh_{j,\,eq}^{g}\,R_{qq}^{e} \\
\end{array}
\label{eq/Measurable/02}
\end{equation}

Using integral relations \eqr{eq/Resistivities/17} for absolute resistivities together with \eqr{eq/Measurable/02} and \eqr{eq/Measurable/06}, we find integral
relations for the measurable resistivities:
\begin{equation}  \label{eq/Measurable/03}
\begin{array}{rl}
R^{\,\prime\,g}_{qq} & = \excess{r_{qq}}
\\\\
R^{\,\prime\,g}_{qi} & = \excess{r_{qi}} + \excess{(\wh_{i,eq}^{g}-\wh_{i,eq})\,r_{qq}}
\\\\
R^{\,\prime\,g}_{ji} & =  \excess{r_{ji}}%
                        + \excess{(\wh_{i,eq}^{g}-\wh_{i,eq})\,r_{jq}} + \excess{(\wh_{j,eq}^{g}-\wh_{j,eq})\,r_{qi}} %
                        + \excess{(\wh_{j,eq}^{g}-\wh_{j,eq})\,(\wh_{i,eq}^{g}-\wh_{i,eq})\,r_{qq}} \\
\end{array}
\end{equation}
For the record we also give integral relations for the measurable interfacial resistances on the liquid side of the surface:
\begin{equation}  \label{eq/Measurable/03a}
\begin{array}{rl}
R^{\,\prime\,\ell}_{qq} & = \excess{r_{qq}}
\\\\
R^{\,\prime\,\ell}_{qi} & = \excess{r_{qi}} + \excess{(\wh_{i,eq}^{\ell}-\wh_{i,eq})\,r_{qq}}
\\\\
R^{\,\prime\,\ell}_{ji} & =  \excess{r_{ji}}%
                        + \excess{(\wh_{i,eq}^{\ell}-\wh_{i,eq})\,r_{jq}} + \excess{(\wh_{j,eq}^{\ell}-\wh_{j,eq})\,r_{qi}} %
                        + \excess{(\wh_{j,eq}^{\ell}-\wh_{j,eq})\,(\wh_{i,eq}^{\ell}-\wh_{i,eq})\,r_{qq}} \\
\end{array}
\end{equation}

Note, that the excess resistances are linear in $\xs$. For example, if we consider the two different positions of the dividing surface $\xs_{1}$ and $\xs_{2}$, then
for the excess resistance to the heat transfer we get
\begin{equation}  \label{eq/Measurable/10}
R^{\,\prime\,\ell}_{qq}(\xs_{1}) - R^{\,\prime\,\ell}_{qq}(\xs_{2}) = (\xs_{1}-\xs_{2})(r_{qq}^{\ell}-r_{qq}^{g})
\end{equation}
The similar relations hold for other resistances.

\subsection{Transport along the interface}\label{sec/Measurable/Parallel}

In contrast to the transport perpendicular to the surface, in the case of the transport parallel to the surface we are not interested in the extrapolated
quantities. The reason for this is, as was mentioned above, that the surface is a separate entity and the transport along the surface cannot be determined by the
properties of adjacent phases. This observation is in agreement with the hypothesis of local equilibrium of the surface, which we verified in \cite{glav/grad2}. In
view of this the measurable heat flux along the surface should be defined differently from the one  normal to the surface. The measurable heat flux is the total
energy flux corrected for the comoving energy due to the mass flux. As the surface is a separate entity, we should substract the excess enthalpy flux of the whole
surface from the excess of the total energy flux.

The enthalpy of the surface is defined using the excess densities. If $\phi(\vR)$ is a specific quantity per unit of mass, the corresponding surface specific
quantity is defined as \cite{bedeaux/advchemphys}
\begin{equation}\label{eq/Excess/Definition/03a}
\phi^{s} \equiv \frac{\excessD{\rho\,\phi}}{\excessD{\rho}}
\end{equation}
where $\rho$ is the mass density. Given that $\phi(\vR)$ has a dimensionality of a quantity per unit of mass, the surface quantity $\phi^{s}$ has a dimensionality
of a quantity per unit of mass as well.

We define the measurable heat flux along the interface as
\begin{equation}\label{eq/ExcessEntropy/03b}
\vJ_{q,\parallel}^{\,\prime,\,s} \equiv \vJ_{e,\parallel}^{s} - \sum_{i=1}^{n}{\wh_{i}^{s}\,\vJ_{\xi_{i},\parallel}^{s}}
\end{equation}
where the surface enthalpies $\wh_{i}^{s}$ are defined using \eqr{eq/Excess/Definition/03a}. We emphasize the difference between the perpendicular and the parallel
transport here. In the case of the transport normal to the surface, we have the measurable fluxes on the gas side of the surface, $J_{q,\perp}^{\,\prime,\,g}$, and
on the liquid side of the surface, $J_{q,\perp}^{\,\prime,\,\ell}$, which are different from each other. In the case of the transport along the surface, we only
have the measurable flux $\vJ_{q,\parallel}^{\,\prime,\,s}$.

In terms of the measurable heat flux the excess entropy production due to transport along the surface \eqref{eq/ExcessEntropy/02a} can be written using
\eqr{eq/ExcessEntropy/03b} to linear order:
\begin{equation}\label{eq/ExcessEntropy/10}
\sigma_{s,\parallel}^{s} = \vJ_{q,\parallel}^{\,\prime,\,s}\spd\nabla_{\parallel}\frac{1}{T^{s}} - \sum_{i=1}^{n}{\vJ_{\xi_{i},\parallel}^{s}
\spd\left[\nabla_{\parallel}\frac{\widetilde{\mu_{i}}^{s}}{T^{s}} - \wh_{i,eq}^{s}\nabla_{\parallel}\frac{1}{T^{s}} \right] }
\end{equation}
where we replaced the surface enthalpies $\wh_{i}^{s}$ with their equilibrium values $\wh_{i,eq}^{s}$ which is correct to linear order.

The corresponding force-flux relations are
\begin{equation}\label{eq/ExcessEntropy/11}
\begin{array}{rl}
\vJ_{q,\parallel}^{\,\prime,\,s} &= \displaystyle L_{qq}^{\,\prime\,s}\nabla_{\parallel}\frac{1}{T^{s}} +
\sum_{i=1}^{n}{L_{qi}^{\,\prime\,s}\left(-\nabla_{\parallel}\frac{\widetilde{\mu_{i}}^{s}}{T^{s}}+\wh_{i,eq}^{s}\nabla_{\parallel}\frac{1}{T^{s}}\right)}
\\\\
\vJ_{\xi_j,\parallel}^{\,\prime,\,s} &= \displaystyle L_{jq}^{\,\prime\,s}\nabla_{\parallel}\frac{1}{T^{s}} +
\sum_{i=1}^{n}{L_{ji}^{\,\prime\,s}\left(-\nabla_{\parallel}\frac{\widetilde{\mu_{i}}^{s}}{T^{s}}+\wh_{i,eq}^{s}\nabla_{\parallel}\frac{1}{T^{s}}\right)}
\end{array}
\end{equation}
where we expressed fluxes in terms of forces, as we did it before for parallel contributions. Comparing \eqr{eq/Resistivities/02a} and \eqr{eq/ExcessEntropy/11} we
get for the measurable conductances
\begin{equation}\label{eq/ExcessEntropy/12}
\begin{array}{rl}
L_{qq}^{\,\prime\,s} &= \displaystyle L_{qq}^{e} - \sum_{i=1}^{n}{\wh_{i,\,eq}^{s}\,(L_{iq}^{e}+L_{qi}^{e})} +
\sum_{i=1}^{n}\sum_{j=1}^{n}{\wh_{i,\,eq}^{s}\,\wh_{j,\,eq}^{s}\,L_{ji}^{e}}
\\\\
L_{qi}^{\,\prime\,s} &= \displaystyle L_{qi}^{e} - \sum_{j=1}^{n}{\wh_{j,\,eq}^{s}\,L_{ji}^{e}}
\\\\
L_{ji}^{\,\prime\,s} &= \displaystyle L_{ji}^{e}
\end{array}
\end{equation}

Using \eqr{eq/Resistivities/27} for absolute conductivities together with \eqr{eq/ExcessEntropy/12} and \eqr{eq/ExcessEntropy/13}, we find integral relations for
measurable conductivities
\begin{equation}  \label{eq/ExcessEntropy/15}
\begin{array}{rl}
L^{\,\prime\,s}_{qq} & = \displaystyle \excess{\ell_{qq}} - \sum_{j=1}^{n}\excess{(\wh_{j,eq}^{s}-\wh_{j,eq})\,(\ell_{jq}+\ell_{qj})} %
+ \sum_{j=1}^{n}\sum_{j=1}^{n}\excess{(\wh_{j,eq}^{s}-\wh_{j,eq})\,(\wh_{i,eq}^{s}-\wh_{i,eq})\,\ell_{qq}}
\\\\
L^{\,\prime\,s}_{qi} & = \displaystyle \excess{\ell_{qi}} - \sum_{j=1}^{n}\excess{(\wh_{j,eq}^{s}-\wh_{j,eq})\,\ell_{ji}}
\\\\
L^{\,\prime\,s}_{ji} & = \displaystyle \excess{\ell_{ji}}  \\
\end{array}
\end{equation}
$L^{\,\prime\,s}_{ji}$ depends linearly on the position of $\xs$, $L^{\,\prime\,s}_{qi}$ depends on $\xs$ quadratically, and $L^{\,\prime\,s}_{qq}$ depends on $\xs$
cubically.

\section{Discussion}\label{sec/Discussion}
\subsection{Dependence on the enthalpy profile}\label{sec/Discussion/Enthalpy}

An important property to analyze is the dependence of the thermodynamic quantities on the reference state. All the thermodynamic potentials are defined with respect
to some reference state. The behavior of the resistances $R^{e}$ and resistivities $r^{e}$ on one hand and $R^{\,\prime\,g}$ and $r^{\,\prime}$ on the other hand is
different in this respect. Both $R^{e}$ and $r^{e}$ coefficients depend on the reference state, as they are associated with the absolute fluxes. The coefficients
$R^{\,\prime\,g}$ and $r^{\,\prime}$ are associated with the measurable fluxes and therefore independent of the reference state. The similar arguments are valid for
the interfacial conductances and conductivities.

Consider integral relations for the measurable resistances, \eqr{eq/Measurable/03} or \eqr{eq/Measurable/03a}. Each component of the resistance matrix
$\mathrm{R}^{\,\prime\,g}$ contains the excess of the local resistivity profiles coupled to the enthalpy profiles. Since all the terms on the right hand side of
these equations are evaluated at equilibrium, the resistances depend linearly on the position of the dividing surface.

The first term in the expression for every resistance is the excess of only the corresponding resistivity profile. As we discussed in \cite{glav/grad1}, each
resistivity profile contains a peak due to the interfacial contribution \cite{surfres}. It is the size of this peak which controls the magnitude of the first term
and makes it positive. This contribution is the only contribution to the interfacial resistance $R_{qq}^{\,\prime,\,g} = R_{qq}^{\,\prime,\,\ell}$ to heat flow. In
this sense, the resistance of the surface to heat flow is similar to the heat resistance of a finite bulk layer.

\begin{figure}
\centering
\subfigure[] %
{\includegraphics[scale=\profilescale]{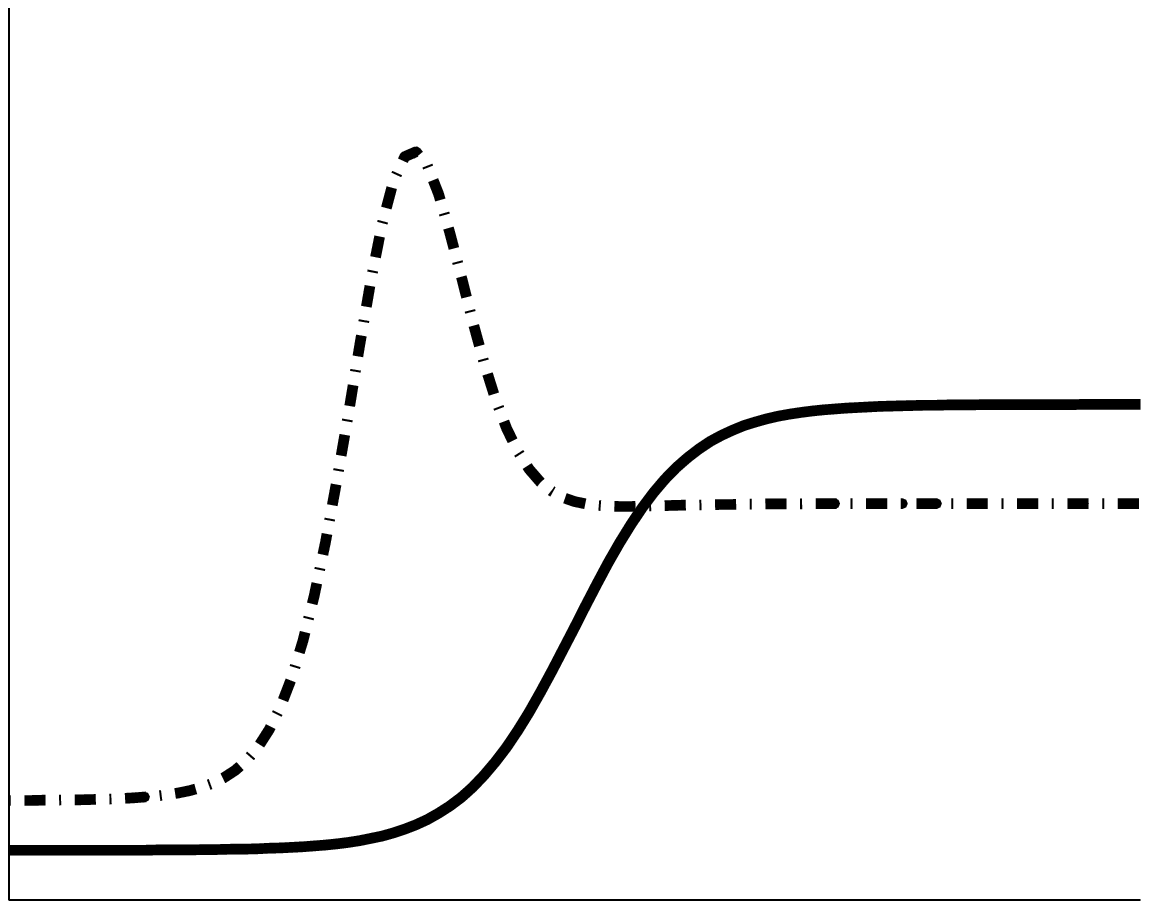}\label{hrprofile1}} %
\subfigure[] %
{\includegraphics[scale=\profilescale]{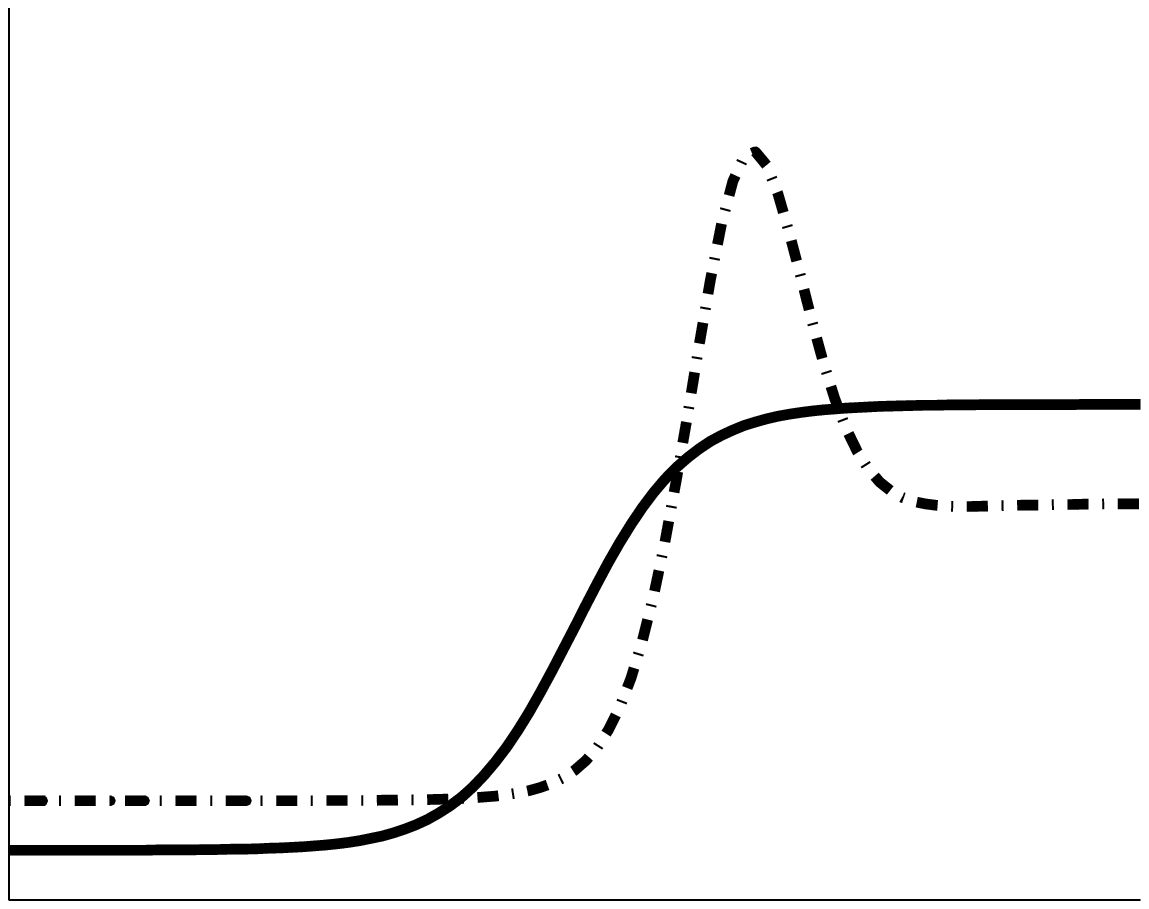} \label{hrprofile2}}  %
\caption{Schematic representation of the enthalpy profile $h_{i}(x)$ (solid line) and the resistivity profile $r_{qq}(x)$ (dash-dotted line): \ref{hrprofile1} the
resistivity profile is shifted to the left; \ref{hrprofile2} the resistivity profile is shifted to the right.}\label{hrprofiles}
\end{figure}

The resistances to the mass transfer due to the temperature difference, $R_{qi}^{\,\prime}$, contain the excess of an additional contributions, however. This
contribution depend on the variation of the enthalpy profile across the interface $\wh_{i,eq}^{g}-\wh_{i,eq}$ and the local heat resistivity profile. The variation
of the enthalpy difference changes from zero on the gas side of the surface to the value of the enthalpy of vaporization $\Delta h_{i}^{vap} \equiv h_{i}^{\ell} -
h_{i}^{g}$ on the liquid side. This gives an enormous contribution, negative for $R_{qi}^{\,\prime,\,g}$ and positive for $R_{qi}^{\,\prime,\,\ell}$, to the
resistances to the mass transfer due to the temperature difference. The magnitude of this contribution depends on the relative position of the enthalpy and the
resistivity profiles. This is illustrated in \figr{hrprofiles}, where the peak in $r_{qq}$ hardly contributes to the integral of the product in \figr{hrprofile1},
while it contributes substantially in \figr{hrprofile2}.
Depending on the sign of the coupling coefficient, the other contributions may be positive or negative.

In contrast to the resistances, the conductances are not linear with respect to the position of the dividing surface. This is due to the dependence on the excess
enthalpy of the surface on the position of the dividing surface. The other arguments given above for the resistances, are similarly applicable to the conductances.

\subsection{Influence of the system geometry}\label{sec/Discussion/Geometry}

Here we consider some typical cases of the surface geometry. As the analysis above is general, we apply it to planar interfaces and droplets.

For planar interface the lame coefficients $\lame_{1} = \lame _{2} = \lame_{3} = 1$ and the excesses defined above in \eqr{eq/Excess/Definition/03},
\eqr{eq/Resistivities/14}, and \eqr{eq/Resistivities/24} take the following form:
\begin{equation}  \label{eq/Geometry/01}
\begin{array}{rl}
\excessD{\phi}(\xs) & = \mathfrak{P}\{\phi\}(\xs) \\
&  \\
\excessR{r}(\xs) & = \mathfrak{P}\{r\}(\xs)\\
&  \\
\excessL{\ell}(\xs) & = \mathfrak{P}\{\ell\}(\xs)
\end{array}
\end{equation}
where
\begin{equation}  \label{eq/Geometry/00}
\mathfrak{P}\{q\} \equiv \displaystyle \int_{\xgsb}^{ \xlsb}{dx\,q^{ex}(x; \xs)}
\end{equation}
is the excess operator for planar interface and $q^{ex}$ is defined similarly to \eqr{eq/Excess/Definition/04}. We see, that for planar surface the approximation
made throughout the paper is satisfied exactly.

For droplets, which have spherical symmetry, we have $\lame_{1} = 1$, $\lame_{2} = x$, $\lame_{3} = x \sin\theta$ where $x$ is the position coordinate and $\theta$
is the polar coordinate. Then, the excesses defined above take the following form:
\begin{equation}  \label{eq/Geometry/02}
\begin{array}{rl}
\excessD{\phi}(\xs) & = \displaystyle \int_{\xgsb}^{\xlsb} {dx\,\left(\frac{x}{\xs}\right)^{2}\,\phi^{ex}(x; \xs)} \\
&  \\
\excessR{r}(\xs) & = \displaystyle \int_{\xgsb}^{\xlsb}{dx\,\left(\frac{\xs}{x}\right)^{2}\,r^{ex}(x; \xs)} \\
&  \\
\excessL{\ell}(\xs) & = \displaystyle \int_{\xgsb}^{\xlsb}{dx\,\ell^{ex}(x; \xs)} \\
\end{array}
\end{equation}
where $\xs$ is the radius of the dividing surface, $\xgsb$ and $\xlsb$ are the radiuses of the surface boundaries.

Assuming that the surface width is small we may expand $x$ around $\xs$. Then the corresponding excesses take the following form
\begin{equation}  \label{eq/Geometry/03}
\begin{array}{rl}
\excessD{\phi}(\xs) & = \displaystyle \mathfrak{P}\{\phi\}(\xs) + \frac{2}{\xs}\,\mathfrak{P}_{1}\{\phi\}(\xs) + \ldots \\
&  \\
\excessR{r}(\xs) & = \displaystyle \mathfrak{P}\{r\}(\xs) - \frac{2}{\xs}\,\mathfrak{P}_{1}\{r\}(\xs) + \ldots \\
&  \\
\excessL{\ell}(\xs) & = \displaystyle \mathfrak{P}\{\ell\}(\xs)
\end{array}
\end{equation}
where $\mathfrak{P}\{q\}$ is the planar excess operator, given by \eqr{eq/Geometry/00}, while $\mathfrak{P}_{1}\{q\}(\xs)$ is the first moment of the planar excess,
given by
\begin{equation}  \label{eq/Geometry/04}
\mathfrak{P}_{1}\{q\}(\xs) \equiv \displaystyle \int_{\xgsb-\xs}^{\xlsb-\xs} {dx\,x\,q^{ex}(x; \xs)}
\end{equation}

We see, that if the radius of the droplet is large, all the excesses may be approximated by the planar interface excesses. As the radius decreases, the
contributions proportional to $1/\xs$ become more and more significant. Particularly for small droplets, the resistance has an important curvature contribution.
Contrary, the interfacial conductance of the droplet does not depend on its size at all. As the spherical droplet has a two-dimensional isotropy, it is also follows
that the excess conductance is still a scalar for a droplet of an arbitrary size.

\section{Conclusions}\label{sec/Conclusions}

In this paper we have build a general approach for the description of the transport phenomena in the interfacial region. We have shown how the continuous
description can be linked to the excess properties of the surface. This makes it possible to consider the surface as an autonomous phase not only in equilibrium but
also in non-equilibrium.

We have derived integral relations for the interfacial resistances for the transport perpendicular to the surface an the interfacial conductances for the transport
parallel to the surface. They are the coefficients in the linear force-flux relations for the heat and mass transfer across and along the interface. The analysis
was done for curved surfaces which allows us to address different applications where the evaporation and condensation happens. In particular, we considered planar
interfaces, which are valuable for distillation processes, and spherical interfaces, which are important in nucleation processes.

Building the link between the continuous and discontinuous description allows one to see the important factors which affect the transport phenomena most. Among them
we can distinguish the local resistivity/conductivity profiles, the enthalpy of vaporization and the curvature of the surface.

The excess of the local resistivity to transport normal to the surface gives the resistance of the whole surface. This is similar to the resistance of a finite bulk
layer, which may be thought as a series of resistors. Similarly, the excess of the local conductivity for transport along the surface gives the conductance of the
whole surface. This is also similar to the conductance of a finite bulk layer, which may be thought of a set of parallel conductors. If there is a peak in the local
profile, it immediately affects the resistance of the conductance of the whole surface. It is therefore crucial to have the information about the whole profile of a
local transport coefficient, not only its bulk values.

Both the interfacial resistances and the interfacial conductances depend on the variation of the enthalpy across the interface. The overall transport coefficients
depend on the equilibrium enthalpies which vary a lot through the interface. One can see from the above formulae, that the dependence on the enthalpy of evaporation
(the difference between the enthalpies of the liquid and gas phases) is crucial not only for the diagonal diffusion coefficient, but also for the off-diagonal
coefficients. This is an important result since cross coefficients are usually neglected in the description of the interfacial phenomena.

The curvature of the surface is important for such processes as nucleation. As the radius of the droplet becomes smaller, the higher moments of the resistivities
contribute to the overall interfacial resistances.

\begin{acknowledgements}
We thank Signe Kjelstrup for valuable comments.
\end{acknowledgements}

\appendix

\section{Excess in curvilinear coordinates}\label{sec/Appendix/Excess/Gradient/3d}

Consider a scalar density $\phi$ being the divergence of a vector function: $ \phi \equiv \nabla\spd\vq(\vR)$. Its excess
\begin{equation}  \label{eq/Appendix/Excess/Gradient/03}
\excessD{\phi}(\xs, \vR_{\parallel}) \equiv \excessD{\nabla\spd\vq}(\xs, \vR_{\parallel}) = \displaystyle \frac{1}{\lame_{2}^{s}\,\lame
_{3}^{s}}\,\int_{\xgsb}^{\xlsb}{dx_{1}\, \lame_{1}\,\lame_{2}\,\lame_{3}\,(\nabla\spd\vq)^{ex}(\vR; \xs)}
\end{equation}
If a density $\phi$ is given by a parallel divergence $\phi \equiv \nabla_{\parallel}\spd\vq_{\parallel}(\vR)$, then its excess is
\begin{equation}  \label{eq/Appendix/Excess/Gradient/03a}
\excessD{\phi}(\xs, \vR_{\parallel}) \equiv \excessD{\nabla_{\parallel}\spd\vq_{\parallel}}(\xs, \vR_{\parallel}) = \displaystyle \frac{1}{\lame_{2}^{s}\,\lame
_{3}^{s}}\,\int_{\xgsb}^{\xlsb}{dx_{1}\, \lame_{1}\,\lame_{2}\,\lame_{3}\,(\nabla_{\parallel}\spd\vq_{\parallel})^{ex}(\vR; \xs)}
\end{equation}

Furthermore, we use the standard formula for the divergence of a vectorial function in curvilinear coordinates
\begin{equation}  \label{eq/Appendix/Excess/Gradient/04}
\nabla\!\cdot\!\mathbf{q} = \displaystyle \frac{1}{\lame_{1}\, \lame_{2}\,\lame_{3}}\left(\frac{\partial}{\partial x_{1}}( \lame_{2}\lame_{3}\,q_{1}) +
\frac{\partial}{\partial x_{2}}( \lame_{1}\lame_{3}\,q_{2}) + \frac{\partial}{\partial x_{3}}( \lame_{1}\lame_{2}\,q_{3}) \right)
\end{equation}
and the parallel divergence of a vectorial function
\begin{equation}  \label{eq/Appendix/Excess/Gradient/04a}
\nabla_{\parallel}\spd\vq = \displaystyle \frac{1}{\lame_{1}\, \lame_{2}\,\lame_{3}}\left(\frac{\partial}{\partial x_{2}}( \lame_{1}\lame_{3}\,q_{2}) +
\frac{\partial}{\partial x_{3}}( \lame_{1}\lame_{2}\,q_{3}) \right)
\end{equation}

According to the definition \eqref{eq/Excess/Definition/04}
\begin{equation}  \label{eq/Appendix/Excess/Gradient/01}
\begin{array}{rl}
(\nabla\spd\vq)^{ex}(\vR; \xs) & = \nabla\spd\vq(\vR) - \{\nabla\spd\vq^{g}(\vR)\}\,\Theta(\xs-x_{1}) - \{\nabla\spd\vq^{\ell}(\vR)\}\,\Theta(x_{1}-\xs) \\
&  \\
& = \nabla\spd(\vq^{ex})(\vR; \xs) + \vq^{g}(\vR)\spd\nabla\Theta(\xs-x_{1}) + \vq^{\ell}(\vR)\spd\nabla\Theta(x_{1}-\xs)
\end{array}
\end{equation}
where $\vq^{ex}(\vR; \xs)$ is defined similarly to \eqr{eq/Excess/Definition/06} and \eqr{eq/Excess/Definition/04}.

Consider $\nabla\spd(\vq^{ex})$, the first term in the second line of \eqr{eq/Appendix/Excess/Gradient/01}. Using \eqr{eq/Appendix/Excess/Gradient/04} one can show
that
\begin{equation}  \label{eq/Appendix/Excess/Gradient/05}
\begin{array}{rl}
\displaystyle \int_{\xgsb}^{\xlsb}{dx_{1}\,\lame_{1}\,\lame_{2}\,\lame_{3}\,\nabla\spd(\vq^{ex})(\vR; \xs)} = %
& \displaystyle \left.\lame_{2}\lame_{3}\,q_{\perp}^{ex}\vphantom{\int}\right|_{\xgsb}^{\xlsb} +\\\\
& \displaystyle + \int_{\xgsb}^{\xlsb}{dx_{1}\left[\frac{\partial}{\partial x_{2}}(\lame_{1}\lame_{3}\,q_{2}^{ex}) + \frac{\partial}{\partial
x_{3}}(\lame_{1}\lame_{2}\,q_{3}^{ex})\right]}
\end{array}
\end{equation}
The first term in \eqr{eq/Appendix/Excess/Gradient/05} vanishes since, according to \eqr{eq/Excess/Definition/01}, $q_{\perp}^{ex}(x^{g,s}) =
q_{\perp}^{ex}(x^{\ell,s}) = 0$. It is easy to verify, that the second term in \eqr{eq/Appendix/Excess/Gradient/05} is equal to
$\lame_{2}^{s}\,\lame_{3}^{s}\;\excessD{\nabla_{\parallel}\spd\vq_{\parallel}}$.

Consider the last two terms in the second line of \eqr{eq/Appendix/Excess/Gradient/01}. Using the standard formula for the gradient of a scalar function in
curvilinear coordinates
\begin{equation}  \label{eq/Appendix/Excess/Gradient/06}
\nabla\varpi = \displaystyle \frac{1}{\lame_{1}}\frac{\partial \varpi }{\partial x_{1}}\,\mathbf{i_{1}} + \frac{1}{\lame_{2}}\frac{\partial \varpi}{\partial
x_{2}}\,\mathbf{i_{2}} + \frac{1}{\lame_{3}}\frac{
\partial \varpi}{\partial x_{3}}\,\mathbf{i_{3}}
\end{equation}
one can show that for Heaviside step function $\Theta$
\begin{equation}  \label{eq/Appendix/Excess/Gradient/07}
\displaystyle \frac{1}{\lame_{2}^{s}\,\lame_{3}^{s}}\,\int_{ \xgsb}^{\xlsb}{dx_{1}\,\lame_{1}\,\lame_{2}\,\lame_{3}\,\vq^{b}(\vR )\spd\nabla\Theta(x_{1}-\xs)} =
\vq^{b}(\xs, \vR_{\parallel})\spd\mathbf{i}_{1} \equiv q_{\perp}^{b}(\xs, \vR_{\parallel})
\end{equation}

\eqr{eq/Appendix/Excess/Gradient/03} then takes the following form
\begin{equation}  \label{eq/Appendix/Excess/Gradient/09}
\excessD{\nabla\spd\vq}(\xs, \vR_{\parallel}) = q_{\perp}^{\ell}(\xs, \vR_{\parallel}) - q_{\perp}^{g}(\xs, \vR_{\parallel}) +
\excessD{\nabla_{\parallel}\spd\vq_{\parallel}}(\xs, \vR_{\parallel})
\end{equation}
We note, that \eqr{eq/Appendix/Excess/Gradient/09} is the exact equation for the general form of the excess function. No assumptions about the curvature magnitude,
like in \eqr{eq/Excess/Definition/05} or \eqr{eq/Excess/Definition/06}, or the nature of the vectorial function $\vq$ were made here.

\section{Excess of a parallel flux for the thin surface}\label{sec/Appendix/Excess/Gradient/2d}

Consider a scalar function $\varpi(x_{\perp}, \vR_{\parallel})$ such that it is independent on the normal coordinate $x_{\perp}$. Furthermore, assume that
corresponding extrapolated functions $\varpi^{g}$ and $\varpi^{\ell}$ obey the same property. It follows therefore that a function $\varpi$ and the extrapolated
function $\varpi^{b}$ have the following form:
\begin{equation}\label{eq/Appendix/Excess/Gradient/22}
\begin{array}{rl}
\varpi(x_{\perp}, \vR_{\parallel}) & = \varpi(\vR_{\parallel}) \\\\
\varpi^{b}(x_{\perp}, \vR_{\parallel}) & = \varpi^{b}(\vR_{\parallel})
\end{array}
\end{equation}
According to \eqr{eq/Excess/Definition/01}, $\varpi(\xsb_{\perp}, \vR_{\parallel}) = \varpi^{b}(\xsb_{\perp}, \vR_{\parallel})$ which leads to
\begin{equation}\label{eq/Appendix/Excess/Gradient/23}
\varpi(\vR_{\parallel}) = \varpi^{b}(\vR_{\parallel})
\end{equation}

Consider the excess $\excessD{\vJ_{\parallel}\spd\nabla_{\parallel}\varpi}$, where the function $\varpi$ satisfies the above relations. Using the definition
\eqref{eq/Appendix/Excess/Gradient/03a}, one can show that
\begin{equation}  \label{eq/Appendix/Excess/Gradient/24}
\excessD{\vJ_{\parallel}\spd\nabla_{\parallel}\varpi} = \frac{1}{\lame_{2}^{s}\,\lame_{3}^{s}}\,\int_{\xgsb}^{\xlsb}{dx_{1}\,\left[
\lame_{1}\lame_{3}\left(J_{2}\,\frac{\partial\varpi}{\partial x_{2}}\right)^{ex} + \lame_{1}\lame_{2}\left(J_{3}\,\frac{\partial\varpi}{\partial x_{3}}\right)^{ex}
\right]}
\end{equation}
It follows from \eqr{eq/Appendix/Excess/Gradient/23} that
\begin{equation}  \label{eq/Appendix/Excess/Gradient/25}
\excessD{\vJ_{\parallel}\spd\nabla_{\parallel}\varpi} = %
\frac{\partial\varpi}{\partial x_{2}}\,\frac{1}{\lame_{2}^{s}\,\lame_{3}^{s}}\,\int_{\xgsb}^{\xlsb}{dx_{1}\,\lame_{1}\lame_{3}J_{2}^{ex} } +
\frac{\partial\varpi}{\partial x_{3}}\,\frac{1}{\lame_{2}^{s}\,\lame_{3}^{s}}\,\int_{\xgsb}^{\xlsb}{dx_{1}\,\lame_{1}\lame_{2}J_{3}^{ex} }
\end{equation}
Using the definition \eqref{eq/Excess/Definition/05}, this can be written as
\begin{equation}  \label{eq/Appendix/Excess/Gradient/26}
\excessD{\vJ_{\parallel}\spd\nabla_{\parallel}\varpi} = \excessD{\vJ_{\parallel}}\spd\nabla_{\parallel}\varpi
\end{equation}
\section{Local resistivities}\label{sec/Appendix/Resistivities}

We need to relate the resistivities $r^{e}$ from \eqr{eq/Resistivities/01} to the resistivities $r^{\,\prime}$ from \eqr{eq/Measurable/04}. This is done by
comparing the coefficients at the same fluxes in these equations. To do this we need to translate the set of fluxes used in \eqr{eq/Measurable/04},
$\{J_{q}^{\,\prime},\,J_{1},\cdots,J_{n-1}\}$, to the set of fluxes used in \eqr{eq/Resistivities/01}, $\{J_{e},\,J_{\xi_{1}},\cdots,J_{\xi_{n}}\}$. This is done
with the help of the relation
\begin{equation}  \label{eq/Appendix/Resistivities/01}
\begin{array}{rl}
J_{i,\perp} = & \displaystyle J_{\xi_{i},\perp} - \xi_{i}\sum_{k=1}^{n}{J_{\xi_{k},\perp}}
\\
J_{q,\perp}^{\,\prime} = & \displaystyle J_{e,\perp} - \sum_{k=1}^{n}{\wh_{k}J_{\xi_{k},\perp}} \\
&
\end{array}
\end{equation}
Substituting $J_{q,\perp}^{\,\prime}$ and $J_{i,\perp}$ into the first of \eqr{eq/Measurable/04} we obtain
\begin{equation}  \label{eq/Appendix/Resistivities/02}
\displaystyle \nabla_{\perp}\frac{1}{T} = \displaystyle r^{\,\prime}_{qq}\,J_{e,\perp}%
+ \sum_{i=1}^{n-1}{J_{\xi_{i},\perp}\Big( r^{\,\prime}_{qi} - r^{\,\prime}_{qq}\wh_{i} - \sum_{k=1}^{n-1}{r^{\,\prime}_{qk}\xi_{k}}\Big)}  %
- J_{\xi_{n},\perp}\Big(r^{\,\prime}_{qq}\wh_{n} + \sum_{k=1}^{n-1}{r^{\,\prime}_{qk}\xi_{k}}\Big)
\end{equation}
Comparing it with the first of \eqr{eq/Resistivities/01} we obtain
\begin{equation}  \label{eq/Appendix/Resistivities/03}
\begin{array}{rl}
r^{e}_{qq} =& \displaystyle r^{\,\prime}_{qq} \\
r^{e}_{qi} =& - \displaystyle r^{\,\prime}_{qq}\wh_{i} - \sum_{k=1}^{n-1}{r^{\,\prime}_{qk}\,\xi_{k}} + r^{\,\prime}_{qi},\quad i=\nmorange \\
r^{e}_{qn} =& - \displaystyle r^{\,\prime}_{qq}\wh_{n} - \sum_{k=1}^{n-1}{r^{\,\prime}_{qk}\,\xi_{k}} \\
\end{array}
\end{equation}
which are the first 3 equations of \eqr{eq/Measurable/06}.

In order to obtain the remaining relations we consider the second of \eqr{eq/Resistivities/01}, which gives
\begin{equation}  \label{eq/Appendix/Resistivities/04}
\begin{array}{rl}
\displaystyle -\sum_{j=1}^{n}{\xi_{j}\,\nabla_{\perp}\frac{\wmu_{j}}{T}} &=%
\displaystyle J_{e,\perp}\sum_{j=1}^{n}{r^{e}_{jq}\xi_{j}} + \sum_{i=1}^{n}{J_{\xi_{i},\perp}\sum_{j=1}^{n}{r^{e}_{ji}\xi_{j}}} \\
\displaystyle \left(-\nabla_{\perp}\frac{\wmu_{j}}{T}\right) - \left(-\nabla_{\perp}\frac{\wmu_{n}}{T}\right) &= %
\displaystyle J_{e,\perp}(r^{e}_{jq}-r^{e}_{nq}) + \sum_{i=1}^{n}{J_{\xi_{i},\perp}(r^{e}_{ji}-r^{e}_{ni})},\quad j=\nmorange
\end{array}
\end{equation}

Furthermore we use \eqr{eq/ExcessEntropy/04a}. In case of the transport in the direction only perpendicular to the surface $\velocity$ can be taken away and
\eqr{eq/ExcessEntropy/04a} becomes
\begin{equation}  \label{eq/Appendix/Resistivities/05}
\sum_{i=1}^{n}{\xi_{i}\left(\nabla\frac{\wmu_{i}}{T} - \widetilde{ h}_{i}\nabla\frac{1}{T}\right)} = 0
\end{equation}
Together with the second of \eqr{eq/Measurable/04} it gives
\begin{equation}  \label{eq/Appendix/Resistivities/06}
\begin{array}{rl}
\displaystyle -\sum_{i=1}^{n}{\xi_{i}\,\nabla_{\perp}\frac{\wmu_{i}}{T}} &= %
\displaystyle -\sum_{i=1}^{n}{\xi_{i}\,\wh_{i}\,\nabla_{\perp}\frac{1}{T}}\\
\displaystyle -\nabla_{\perp}\frac{\mu_{jn}}{T} &= %
\displaystyle -h_{jn}\nabla_{\perp}\frac{1}{T} + r^{\,\prime}_{jq}\,J_{q}^{\,\prime} + \sum_{i=1}^{n-1}{r^{\,\prime}_{ji}\,J_{i}}
\end{array}
\end{equation}
Substituting $\nabla(1/T)$ from \eqr{eq/Appendix/Resistivities/02} and $J_{q}^{\,\prime}$ and $J_{i}$ from \eqr{eq/Appendix/Resistivities/01} we obtain the left
hand size of \eqr{eq/Appendix/Resistivities/06} expressed in terms of the fluxes $J_{e}$ and $J_{\xi_{1}}$ and the resistivities $r^{\,\prime}$. Comparing the
result with \eqr{eq/Appendix/Resistivities/04} we obtain the following equations sets
\begin{subequations}\label{eq/Appendix/Resistivities/07}
\begin{equation}  \label{eq/Appendix/Resistivities/07a}
\begin{array}{rl}
\sum_{k=1}^{n}{r^{e}_{kq}\xi_{k}}  &= - r^{\,\prime}_{qq}\sum_{k=1}^{n}{\xi_{k}\,\wh_{k}}\\\\
r^{e}_{jq} - r^{e}_{nq} &=  - r^{\,\prime}_{qq}h_{jn} + r^{\,\prime}_{jq}, \quad j=\nmorange \\
\end{array}
\end{equation}
\begin{equation}  \label{eq/Appendix/Resistivities/07b}
\begin{array}{rl}
\sum_{k=1}^{n}{r^{e}_{kj}\xi_{k}}   =& (r^{\,\prime}_{qq}\wh_{i} + \sum_{k=1}^{n-1}{r^{\,\prime}_{qk}\,\xi_{k}} - r^{\,\prime}_{qi}) \sum_{k=1}^{n}{\xi_{k}\,\wh_{k}}\\\\
r^{e}_{ji} - r^{e}_{ni}             =& (r^{\,\prime}_{qq}\wh_{i} + \sum_{k=1}^{n-1}{r^{\,\prime}_{qk}\,\xi_{k}} - r^{\,\prime}_{qi})\,h_{jn} \\\\
                                    & - \sum_{k=1}^{n-1}{r^{\,\prime}_{jk}\,\xi_{k}} - r^{\,\prime}_{jq}\wh_{i} + r^{\,\prime}_{ji}, \quad j,i=\nmorange\\
\end{array}
\end{equation}
\begin{equation}  \label{eq/Appendix/Resistivities/07c}
\begin{array}{rl}
\sum_{k=1}^{n}{r^{e}_{kn}\xi_{k}}  =& (r^{\,\prime}_{qq}\wh_{n} + \sum_{k=1}^{n-1}{r^{\,\prime}_{qk}\,\xi_{k}}) \sum_{k=1}^{n}{\xi_{k}\,\wh_{k}}\\\\
r^{e}_{jn} - r^{e}_{nn}            =& (r^{\,\prime}_{qq}\wh_{n} + \sum_{k=1}^{n-1}{r^{\,\prime}_{qk}\,\xi_{k}})\,h_{jn} \\\\
                                    & -\sum_{k=1}^{n-1}{r^{\,\prime}_{jk}\,\xi_{k}} - r^{\,\prime}_{jq}\wh_{n}, \quad j=\nmorange\\
\end{array}
\end{equation}
\end{subequations}
solving which we obtain the relations between the remaining resistivities
\begin{equation}  \label{eq/Appendix/Resistivities/08}
\begin{array}{rl}
r^{e}_{jq} =& \displaystyle -r^{\,\prime}_{qq}\wh_{j} - \sum_{k=1}^{n-1}{r^{\,\prime}_{kq}\,\xi_{k}} + r^{\,\prime}_{jq},\quad j=\nmorange\\
r^{e}_{nq} =& \displaystyle -r^{\,\prime}_{qq}\wh_{n} - \sum_{k=1}^{n-1}{r^{\,\prime}_{kq}\,\xi_{k}} \\
r^{e}_{ji} =& \displaystyle r^{\,\prime}_{qq}\wh_{j}\wh_{i} + \sum_{k=1}^{n-1}{\xi_{k}(r^{\,\prime}_{kq}\wh_{i}+r^{\,\prime}_{qk}\wh_{j})} -(r^{\,\prime}_{jq}\wh_{i}+r^{\,\prime}_{qi}\wh_{j}) \\
            & \displaystyle + \sum_{k=1}^{n-1}\sum_{l=1}^{n-1}{r^{\,\prime}_{kl}\,\xi_{k}\,\xi_{l}} - \sum_{k=1}^{n-1}{\xi_{k}\,(r^{\,\prime}_{ki} + r^{\,\prime}_{jk})} + r^{\,\prime}_{ji} , \quad j,i=\nmorange\\
r^{e}_{jn} =& \displaystyle r^{\,\prime}_{qq}\wh_{j}\wh_{n} + \sum_{k=1}^{n-1}{\xi_{k}(r^{\,\prime}_{kq}\wh_{n}+r^{\,\prime}_{qk}\wh_{j})} - r^{\,\prime}_{jq}\wh_{n} \\
            & \displaystyle + \sum_{k=1}^{n-1}\sum_{l=1}^{n-1}{r^{\,\prime}_{kl}\,\xi_{k}\,\xi_{l}} - \sum_{k=1}^{n-1}{r^{\,\prime}_{jk}\,\xi_{k}}, \quad j=\nmorange\\
r^{e}_{ni} =& \displaystyle r^{\,\prime}_{qq}\wh_{n}\widetilde{h} _{i} + \sum_{k=1}^{n-1}{\xi_{k}(r^{\,\prime}_{kq}\wh_{i}+r^{\,\prime}_{qk}\wh_{n})} - r^{\,\prime}_{qi}\wh_{n} \\
            & \displaystyle  + \sum_{k=1}^{n-1}\sum_{l=1}^{n-1}{r^{\,\prime}_{kl}\,\xi_{k}\,\xi_{l}} - \sum_{k=1}^{n-1}{r^{\,\prime}_{ki}\,\xi_{k}}, \quad i=\nmorange \\
r^{e}_{nn} =& \displaystyle r^{\,\prime}_{qq}\wh_{n}^{2}  + \wh_{n}\sum_{k=1}^{n-1}{\xi_{k}(r^{\,\prime}_{kq}+r^{\,\prime}_{qk})} + \sum_{k=1}^{n-1}\sum_{l=1}^{n-1}{r^{\,\prime}_{kl}\,\xi_{k}\,\xi_{l}} \\
\end{array}
\end{equation}

As one can confirm the symmetry of the $r^{\,\prime}$-matrix leads to the symmetry of the $r^{e}$-matrix and vice versa. We therefore do not give the expressions
for $r^{e}_{jq}$, $r^{e}_{nq}$ and $r^{e}_{jn}$ in \eqr{eq/Measurable/06}.

\section{Local conductivities}\label{sec/Appendix/Conductivities}

We need to relate the conductivities $\ell^{e}$ from \eqr{eq/Resistivities/01a} to the conductivities $\ell^{\,\prime}$ from  \eqr{eq/Measurable/04a}. One can do
that by comparing the coefficients at the same forces in those equations, like it was done for resistivities for the transport normal to the surface (see
\appr{sec/Appendix/Resistivities}). We will employ a different procedure for conductivities, however. We will compare the coefficients at the same terms in the
expression for the entropy production. Both procedures use the same method, so they are equivalent. Namely they use the fact that if expressions
$\sum\alpha_{i}x_{i} = \sum\beta_{i}x_{i}$ for linearly independent set $\{x_{i}\}$, then $\alpha_{i} = \beta_{i}$. Note, that even though the size of the
conductivity (resistivity) matrix is different for the two sets we are going to relate, all the fluxes and forces in the corresponding set are linearly independent.
Thus, both $n$ forces $\left\{\nabla(1/T), -\nabla(\mu_{1n}/T)+h_{1n}\nabla(1/T), \cdots, -\nabla(\mu_{n-1,n}/T)+h_{n-1,n}\nabla(1/T) \right\}$  are linearly
independent, as well as $n+1$ forces $\left\{\nabla(1/T), -\nabla(\mu_{1}/T), \cdots, -\nabla(\mu_{n}/T)\right\}$ are linearly independent (the similar statement is
true for fluxes in the case of normal transport through the interface).

We use the equivalent expressions \eqref{eq/ExcessEntropy/05b} and \eqref{eq/Entropy/Balance/05b} for the local entropy production in the parallel direction.

We first substitute \eqr{eq/Measurable/04a} into \eqr{eq/ExcessEntropy/05b} in order to obtain the expression for the entropy production in terms of thermodynamic
forces only, $\nabla(1/T)$ and $-\nabla(\mu_{in}/T)+h_{in}\nabla(1/T)$ for $i=\nmorange$. We then substitute $\mu_{in} = \wmu_{i}-\wmu_{n}$ and write this entropy
production in terms of the the forces $\nabla(1/T)$ and $-\nabla(\mu_{in}/T)$ for $i=\nrange$. After some algebra we obtain the following expression for the entropy
production
\begin{equation}  \label{eq/Appendix/Conductivities/01}
\begin{array}{rl}
\sigma_{s,\parallel} =& %
\displaystyle \left|\nabla_{\parallel}{\frac{1}{T}}\right|^{2} \left[ \ell_{qq}^{\,\prime}%
+ \sum_{i=1}^{n-1}{h_{in}\,(\ell_{iq}^{\,\prime} + \ell_{qi}^{\,\prime})} %
+ \sum_{i=1}^{n-1}\sum_{j=1}^{n-1}{h_{in}h_{jn}\,\ell_{ij}^{\,\prime}} %
\right]
\\\\
&\displaystyle + \sum_{k=1}^{n-1}{\left(-\nabla_{\parallel}\frac{\wmu_{k}}{T}\right)\spd\left(\nabla\frac{1}{T}\right)} \left[ %
\ell_{qk}^{\,\prime} + \ell_{kq}^{\,\prime} +  \sum_{i=1}^{n-1}{h_{in}(\ell_{ik}^{\,\prime}+\ell_{ki}^{\,\prime})}%
\right]
\\\\
&\displaystyle \qquad - {\left(-\nabla_{\parallel}\frac{\wmu_{n}}{T}\right)\spd\left(\nabla\frac{1}{T}\right)} \left[ %
\sum_{k=1}^{n-1}{(\ell_{qk}^{\,\prime} + \ell_{kq}^{\,\prime})} + \sum_{k=1}^{n-1}\sum_{i=1}^{n-1}{h_{in}(\ell_{ik}^{\,\prime}+\ell_{ki}^{\,\prime})}%
\right]
\\\\
&\displaystyle
+\sum_{i=1}^{n-1}\sum_{j=1}^{n-1}{\left(-\nabla_{\parallel}\frac{\wmu_{i}}{T}\right)\spd\left(-\nabla_{\parallel}\frac{\wmu_{j}}{T}\right)} \left[%
\SumVarea \ell_{ji}^{\,\prime}%
\right]
\\\\
&\displaystyle \qquad -\left(-\nabla_{\parallel}\frac{\wmu_{n}}{T}\right)\spd\sum_{k=1}^{n-1}{\left(-\nabla_{\parallel}\frac{\wmu_{k}}{T}\right)} \left[%
\sum_{i=1}^{n-1}{(\ell_{ik}^{\,\prime}+\ell_{ki}^{\,\prime})}%
\right]
\\\\
&\displaystyle \qquad + \left(-\nabla_{\parallel}\frac{\wmu_{n}}{T}\right)\spd\left(-\nabla_{\parallel}\frac{\wmu_{n}}{T}\right)\left[%
\sum_{i=1}^{n-1}\sum_{j=1}^{n-1}{\ell_{ji}^{\,\prime}}%
\right]
\end{array}
\end{equation}

We then substitute  \eqr{eq/Resistivities/01a} into \eqref{eq/Entropy/Balance/05b} and obtain the expression for the entropy production in terms of the
thermodynamic forces $\nabla(1/T)$ and $-\nabla(\mu_{in}/T)$ for $i=\nrange$. As a result we get
\begin{equation}  \label{eq/Appendix/Conductivities/02}
\sigma_{s,\parallel} = %
\left|\nabla_{\parallel}{\frac{1}{T}}\right|^{2} \left[\SumVarea \ell_{qq}^{e} \right] +
\sum_{k=1}^{n}{\left(-\nabla_{\parallel}\frac{\wmu_{k}}{T}\right)\spd\left(\nabla\frac{1}{T}\right)} \left[\SumVarea \ell_{qk}^{e} + \ell_{kq}^{e} \right] +
\sum_{i=1}^{n}\sum_{j=1}^{n}{\left(-\nabla_{\parallel}\frac{\wmu_{i}}{T}\right)\spd\left(-\nabla_{\parallel}\frac{\wmu_{j}}{T}\right)}
\left[\SumVarea\ell_{ji}^{e}\right]
\end{equation}

Comparing the coefficients at corresponding terms in \eqr{eq/Appendix/Conductivities/01} and \eqr{eq/Appendix/Conductivities/02} (in square brackets) we get
\begin{equation}  \label{eq/Appendix/Conductivities/03}
\begin{array}{rl}
\ell_{qq}^{e} =& \displaystyle \ell_{qq}^{\,\prime} %
+ \sum_{i=1}^{n-1}{h_{in}\,(\ell_{iq}^{\,\prime} + \ell_{qi}^{\,\prime})}%
+ \sum_{i=1}^{n-1}\sum_{j=1}^{n-1}{h_{in}h_{jn}\,\ell_{ij}^{\,\prime}} %
\\\\
\ell_{qk}^{e} + \ell_{kq}^{e} = &\displaystyle \ell_{qk}^{\,\prime} + \ell_{kq}^{\,\prime}%
+ \sum_{i=1}^{n-1}{h_{in}(\ell_{ik}^{\,\prime}+\ell_{ki}^{\,\prime})} , \quad k=\nmorange%
\\\\
\ell_{qn}^{e} + \ell_{nq}^{e} = &\displaystyle -\sum_{k=1}^{n-1}{(\ell_{qk}^{\,\prime} + \ell_{kq}^{\,\prime})}%
- \sum_{k=1}^{n-1}\sum_{i=1}^{n-1}{h_{in}(\ell_{ik}^{\,\prime}+\ell_{ki}^{\,\prime})}%
\\\\
\ell_{ji}^{e} = &\displaystyle \ell_{ji}^{\,\prime} , \quad i,j=\nmorange%
\\\\
\ell_{kn}^{e} + \ell_{nk}^{e} = &\displaystyle -\sum_{i=1}^{n-1}{(\ell_{ik}^{\,\prime}+\ell_{ki}^{\,\prime})}, \quad k=\nmorange%
\\\\
\ell_{nn}^{e} = &\displaystyle \sum_{i=1}^{n-1}\sum_{j=1}^{n-1}{\ell_{ji}^{\,\prime}}%
\end{array}
\end{equation}
Without restrictions in generality we assume that the matrix of the coefficients is symmetric, so \eqr{eq/Appendix/Conductivities/03} can be written as
\begin{equation}  \label{eq/Appendix/Conductivities/04}
\begin{array}{rl}
\ell_{qq}^{e} =& \displaystyle \ell_{qq}^{\,\prime} %
+ \sum_{i=1}^{n-1}{h_{in}\,(\ell_{iq}^{\,\prime} + \ell_{qi}^{\,\prime})}%
+ \sum_{i=1}^{n-1}\sum_{j=1}^{n-1}{h_{in}h_{jn}\,\ell_{ij}^{\,\prime}} %
\\\\
\ell_{qk}^{e} = &\displaystyle \ell_{qk}^{\,\prime} + \sum_{i=1}^{n-1}{h_{in}\ell_{ki}^{\,\prime}} , \quad k=\nmorange%
\\\\
\ell_{qn}^{e} = &\displaystyle -\sum_{k=1}^{n-1}{\ell_{qk}^{\,\prime}} - \sum_{k=1}^{n-1}\sum_{i=1}^{n-1}{h_{in}\ell_{ik}^{\,\prime}}%
\\\\
\ell_{ji}^{e} = &\displaystyle \ell_{ji}^{\,\prime} , \quad i,j=\nmorange%
\\\\
\ell_{kn}^{e} = &\displaystyle -\sum_{i=1}^{n-1}{\ell_{ki}^{\,\prime}}, \quad k=\nmorange%
\\\\
\ell_{nn}^{e} = &\displaystyle \sum_{i=1}^{n-1}\sum_{j=1}^{n-1}{\ell_{ji}^{\,\prime}}%
\end{array}
\end{equation}
which leads to \eqr{eq/ExcessEntropy/13} together with \eqr{eq/ExcessEntropy/14}.

\bibliographystyle{unsrt}

\end{document}